\documentclass[useAMS,usenatbib]{mn2e}
\usepackage{rotating}
\usepackage{journals}

\def\gx {GX\,1+4\,\,}

\newcommand{\BSAX}{{\em Beppo}SAX\,\,}

\input psfig.sty

\title[GX 1+4 : A Compton Reflection Dominated Event]{A Compton Reflection Dominated Spectrum In A Peculiar Accreting Neutron Star}
\author[N. Rea et al.]
{N. Rea$^{1,2,3}$, L. Stella$^{1}$, G.L. Israel$^{1}$, G. Matt$^{4}$, S. Zane$^{5}$,  A. Segreto$^{6}$, \newauthor T. Oosterbroek$^{7}$, M. Orlandini$^{8}$ \\
$^{1}$INAF-Osservatorio Astronomico di Roma, via Frascati 33, 00040 Monteporzio
Catone (Roma), Italy \\
$^{2}$Universit\'a degli Studi di Roma ``Tor Vergata'', via della Ricerca Scientifica 1, 00133 Rome, Italy \\
$^{3}$ SRON - Netherlands Institute for Space Research, Sorbonnelaan 2, 3584 CA Utrecht, The Netherlands \\
$^{4}$Universit\'a degli Studi Roma Tre, via della Vasca Navale 84, 00146 Rome,
Italy \\
$^{5}$Mullard Space Science Laboratory, University College London, Holmbury St.
Mary, Dorking, Surrey RH5 6NT \\
$^{6}$INAF-Istituto di Astrofisica Spaziale e Fisica Cosmica - Sezione di Palermo, CNR, via Ugo La Malfa 153, 90146 Palermo, Italy \\
$^{7}$Science Payload and Advanced Concepts Office, ESA--ESTEC, Postbus 299, NL-2200 AG Noordwijk, The Netherlands \\
$^{8}$ INAF-Istituto di Astrofisica Spaziale e Fisica Cosmica - Sezione di Bologna, CNR, via Gobetti 101, I-40129 Bologna, Italy }

\begin{document}

\date{Accepted...Received }
\pagerange{\pageref{firstpage}--\pageref{lastpage}} \pubyear{2005}
\maketitle
\label{firstpage}

\begin{abstract}
  
We report on a puzzling event occurred during a long \BSAX\
observation of the slow-rotating binary pulsar \gx. During this event,
lasting about 1 day, the source X-ray flux was over a factor 10 lower
than normal. The low-energy pulsations disappeared while at higher
energies they were shifted in phase by $\sim 0.25$\,.  The continuum
spectrum taken outside this low-intensity event was well fitted by an
absorbed cut-off power law, and exhibited a broad iron line at
$\sim6.5$\,keV probably due to the blending of the neutral (6.4\,keV)
and ionised (6.7\,keV) $K_{\alpha}$ iron lines. The spectrum during
the event was Compton reflection dominated and it showed two narrow
iron lines at $\sim6.4$\,keV and $\sim7.0$\,keV, the latter never
revealed before in this source. We also present a possible model for
this event in which a variation of the accretion rate thickens a
torus-like accretion disc which hides for a while the direct neutron
star emission from our line of sight.  In this scenario the Compton
reflected emission observed during the event is well explained in
terms of emission reflected by the side of the torus facing our line
of sight.

\end{abstract}


\section{Introduction}

Accreting X-ray pulsars are neutron stars (NSs) in binary systems, the
emission of which is powered by accretion of matter from the
companion. In several of these systems a highly variable X-ray flux is
revealed, usually caused by variations of the mass accretion rate or
by the occurrence of eclipses or dips. 

NS X-ray binaries are usually divided in two sub-classes depending on
the companion mass: namely, the high-mass and low-mass X-ray binaries
(HMXRBs and LMXRBs; for a review see Joss \& Rappaport
1984). Pulsation periods extend over a wide range
($\sim$1.5\,ms--8000\,s); most of the X-ray emission detected from
binary pulsars is in the 2-20\,keV energy range; X-ray spectra are
complex and often cannot be described in terms of a single component.
A usual spectral description is terms of a two-component model,
consisting of a soft component, usually a blackbody or a
disc-blackbody model ($kT\sim0.5-3$ keV), and a hard component
modelled by a power law with a high energy cut-off. There is evidence
in several cases for one or more lines between 6 and 7\,keV, most probably due to iron (Fe) (Becker et al. 1978; White et al. 1980).

Compton reflection of X-rays, has been studied in detail in the last
few decades (see Rybicki \& Lightman 1979; White, Lightman \&
Zdziarski 1988, Lightman \& White 1988) and it might occur when X-ray
and $\gamma$-ray radiation impinges upon a slab of cold material. The
expected UV to X-ray spectrum will then be composed by three different
components: i) the optically thick UV radiation from the thermal
matter (usually modelled by a blackbody), ii) the (possibly
non-thermal) primary X-ray radiation (usually modelled by a power
law), iii) and the reprocessed component by the cold material. The
X-ray spectrum of the latter has a characteristic shape (dictated by
the photoelectric absorption at low energies and Compton
scattering at higher energies), with a broad hump around 30 keV (e.g. Matt et
al. 1991). Many fluorescent emission lines are also present, by far
the most prominent being the iron K$\alpha$ at 6.4 keV. This Compton
reflection component was first observed in Active Galactic Nuclei
(Pounds et al. 1990; Nandra \& Pounds 1994), where it is almost
ubiquitous (Perola et al. 2002; Bianchi et al. 2004), and it is often
present also in X-ray binaries, especially in black hole systems (see
e.g. Done 2004 for a review).

%%%%%%%%%%% FIGURE 1%%%%%%%%%%%%%%%%%%%%%%%%%%%%%

\begin{figure}
\psfig{figure=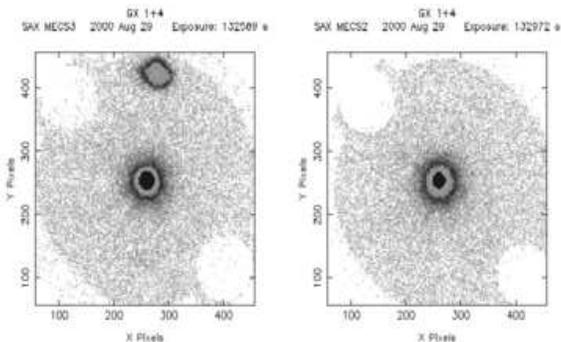,width=8cm,height=5cm,angle=270} 
\caption{Comparison between the MECS3 and MECS2 observations of \gx (1.65--10\,keV energy range). 
A ghost source is present in the MECS3.}\label{comparing}
\label{ghost}
\end{figure}

%%%%%%%%%%%%%%%%%%%%%%%%%%%%%%%%%%%%%%%%%

\gx is a LMXRB system harbouring a $\sim$130\,s pulsar 
(Lewin, Ricker \& McClintock 1971) accreting mass from a red giant
companion (V2116 Ophiuchi) of class M5 III (Davidsen et al. 1977;
Jablonski et al. 1997; Chakrabarty et al. 1997a and 1998; Pereira et
al. 1996). In the NS X-ray binary zoo, we know only another system
with a red giant companion: 4U 1700+24 (Masetti et al. 2002; Galloway
et al. 2002).

\gx shows a variable X-ray flux on virtually all timescales so far 
investigated, from minutes to years. It is a relatively bright X-ray
pulsar system, and it has displayed the largest spin-up rate recorded
for an X-ray pulsar (Lewin, Ricker \& McClintock 1971).  The average
spin-up trend reversed in 1983 to spin-down at approximately the same
rate, since then other changes in the sign of the torque has been
observed for this source (Chakrabarty et al. 1997b). It is somehow a
peculiar object among the X-ray binaries because of the high magnetic
field that the NS is believed to have, which was inferred from its
timing properties ($\sim2-3\times10^{13}$G; Dotani et al. 1989;
Greenhill et al. 1993; Cui 1997). In fact, the presence of such a
high magnetic field in a slowly rotating NS with a red giant
companion is an intriguing puzzle for the evolutionary scenario of
this binary system.
\gx lies in the Galactic plane in the direction of the bulge (RA
17:32:03.0; DEC -24:44:44.3) and it has an atypical spectrum compared
to other X-ray binaries (Frontera \& Dal Fiume 1989). This source has
not shown so far a soft emission component (although it might be
undetected due to the very high absorption, $N_{H} > 2\times10^{22}$
atoms\,cm$^{-2}$); it has a very hard X-ray spectrum which seems to be
completely non-thermal and is usually fitted by a Comptonization model
or a cut-off power law. The spectral parameters are rather variable.
Moreover, a pronounced Fe emission line at $\sim$6.5\,keV is often observed.

In this paper, we report on a long (about 3.5 days) \BSAX observation of 
\gx performed around November 2000. The source showed a marked 
intensity drop event, lasting for about 90\,ks, followed by a recovery
to an almost normal intensity state. We analyzed spectral and timing
variations correlated with this event, and present a model to
interpret the source behaviour.

%%% %%%%% FIGURE 2 %%%%%%%%%%%%

\begin{figure}
\vbox{
\psfig{figure=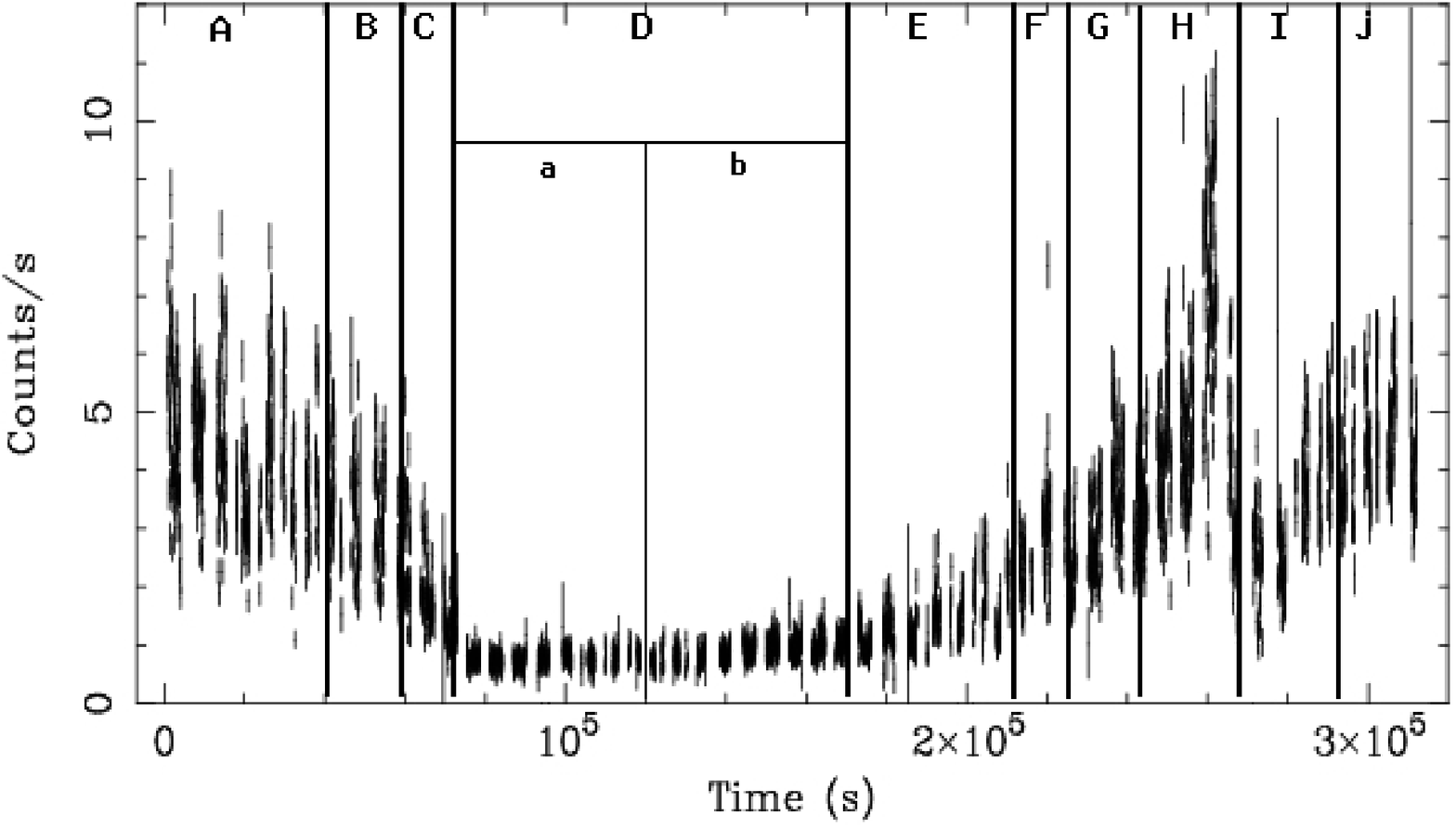,width=8cm,height=5cm} 
\psfig{figure=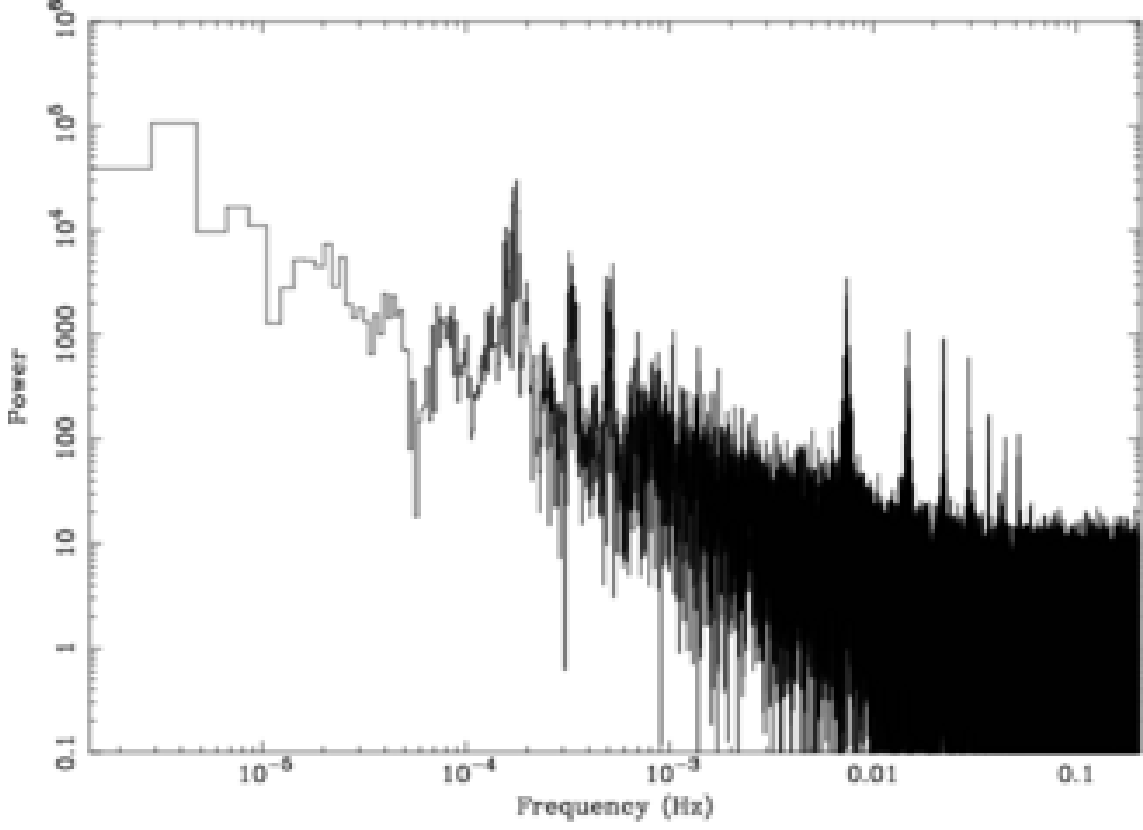,width=8cm,height=4cm,angle=270} 
}
\caption{{\em Top panel}: MECS (1.65--10\,keV) lightcurve of 
the {\em Beppo}SAX 2000 observation (100\,s bins) with the time
interval divisions. {\em Bottom panel}: power spectrum density in the
same energy band. First three peaks are spurious, since due to the
{\em Beppo}SAX orbit, while the others are the NS spin fundamental
peak at $\sim$0.0075 Hz followed by six harmonics.}
\label{lcurve_psd}
\end{figure}

%%%%%%%%%%%%%%%%%%%%%%%%%%

\section{OBSERVATION}

\BSAX\ observed \gx three times, in 1996, in 1997 and in 2000.  The
first two observations are reported elsewhere (Naik et al. 2004),
while here we report on the $\sim$3.5 days observation carried out
between October 29th and November 2nd 2000.  The \BSAX observatory
covered more than three decades of energy, from 0.1 to 200\,keV. The
payload was composed by two Wild Field Cameras (WFC; Jager et al. 1997)
and four co-aligned instrument: the Narrow Field Instruments (NFI:
Boella et al. 1997a; LECS, 0.1-10\,keV, Parmar et al. 1997; MECS,
1-10\,keV, Boella et al. 1997b; HPGSPC, 4-100\,keV, Manzo et al. 1997;
PDS, 15-200\,keV, Frontera et al. 1997).  All four NFI instruments
were on during the observation.

Since the LECS and MECS have imaging capabilities, we extracted the
events from circular regions of 6$^{\prime}$ radii centred on the peak 
of the source point spread function (PSF). 
The LECS background was subtracted using a background
appropriate for low-latitude sources, i.e. source-free observations
close to the galactic plane with a total exposure time of 210 ks (in
order to avoid underestimating the background, we did not use standard
background subtraction from a region in the same image, far from the
source; note that for highly absorbed and intense sources the
low energy background should be subtracted as reported in Parmar et
al. 1999). The MECS background was extracted from an annulus around
the source. The HPGSPC and PDS do not have imaging capabilities so the
background subtraction was obtained using off-source data collected
during the rocking of the collimators.

We carried out timing and spectral analysis using the data collected
from these four instruments. We corrected all arrival times to the
barycenter of the Solar System.

LECS and MECS spectra were accumulated from the same circular regions
used for the event files and re-binned in order to have at least 50
photons per bin; HPGSPC and PDS spectra were re-binned so as to have
at least 80 and 100 photons in each bin, respectively.  With this
choice, the minimum chi-square techniques could be reliably used in
spectral fitting. For the spectral analysis we restricted the energy
range of the instruments to: 1-4\,keV for the LECS (due to the very
high background and absorption value), 1.65-10.5\,keV for the MECS,
9.8-20\,keV for the HPGSPC and 15-200\,keV for the PDS. Only those
bins in which the count rate (after background subtraction) was
significantly higher than zero were used in the spectral analysis. In
time intervals in which the source was dimmest the LECS spectra
contained no useful information, and were thus excluded from the analysis.

LECS response matrices were made using the LECS matrix generation tool
LEMAT (included in SAXDAS 2.2.1) in order to account for the high
count rate of the source (for details see the \BSAX cookbook:
Guainazzi, Fiore \& Grandi 1999). The other instrument's matrices 
are those provided by the \BSAX  data analysis center. 

During the analysis we discovered the presence of a ghost source in
the third MECS instrument (see Fig.\ref{ghost}). The ghost is probably
due to a spurious reflection in the telescope mirrors of a source out
of the instrument field of view. The presence of this source does not
affect the analysis of \gx because it lies $\sim$25' off-axis, and
it is present only in the third MECS. However, given the no--imaging 
capability of
the PDS and the HPGSPC, and their larger field of view, we had to
ensure that photons from this source did not contaminate the high
energy emission of our target. To this aim, we extracted the MECS 3
spectrum of the ghost source, generated an appropriate off-axis
response matrix and fitted its spectrum. Based on the 1--10\,keV
spectrum of this source, we estimated that its 10--100\,keV flux was
more than two orders of magnitude lower than that of \gx in the same
energy band. We concluded that the HPGSPC and the PDS spectra are
largely dominated by our target.

\section{RESULTS}

\subsection{Timing analysis}

In all the energy bands we investigated, the X-ray light-curve 
of the source showed a large flux variability (see
Fig.\ref{lcurve_psd} top panel). In the MECS energy range the source
intensity dramatically dropped from $\sim$4 counts/s to $\sim$0.3
counts/s, then, after a $\sim$90\,ks long low emission state, slowly
increased reaching $\sim$10 counts/s, dropped again for less than
10\,s and then returned to its starting intensity level.

%%%%%%%%%%% FIGURE 3 %%%%%%%%%%%%%%%%%%%%%%%%

\begin{figure*}
\hbox{
\psfig{figure=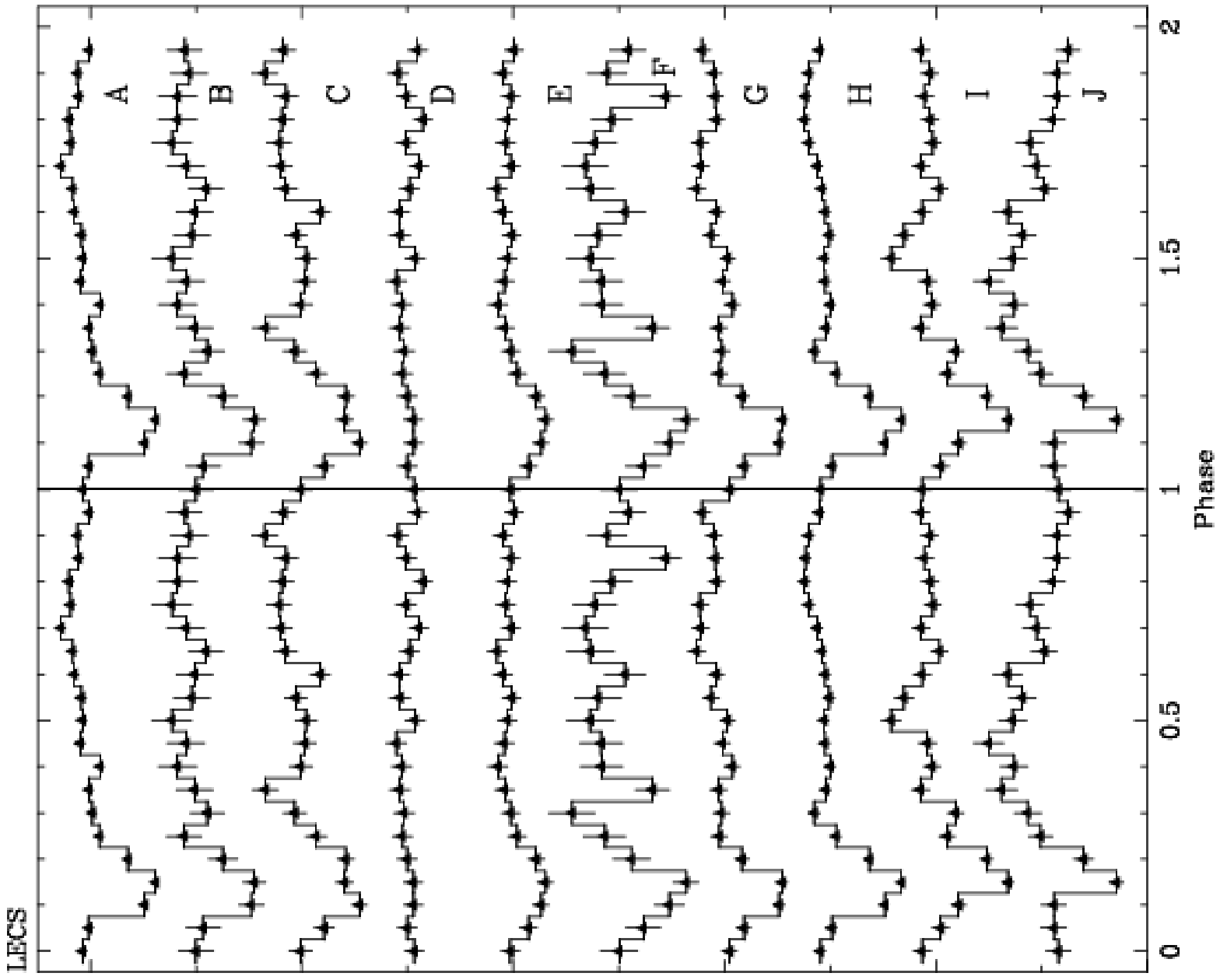,width=4cm,height=12cm,angle=270} 
\psfig{figure=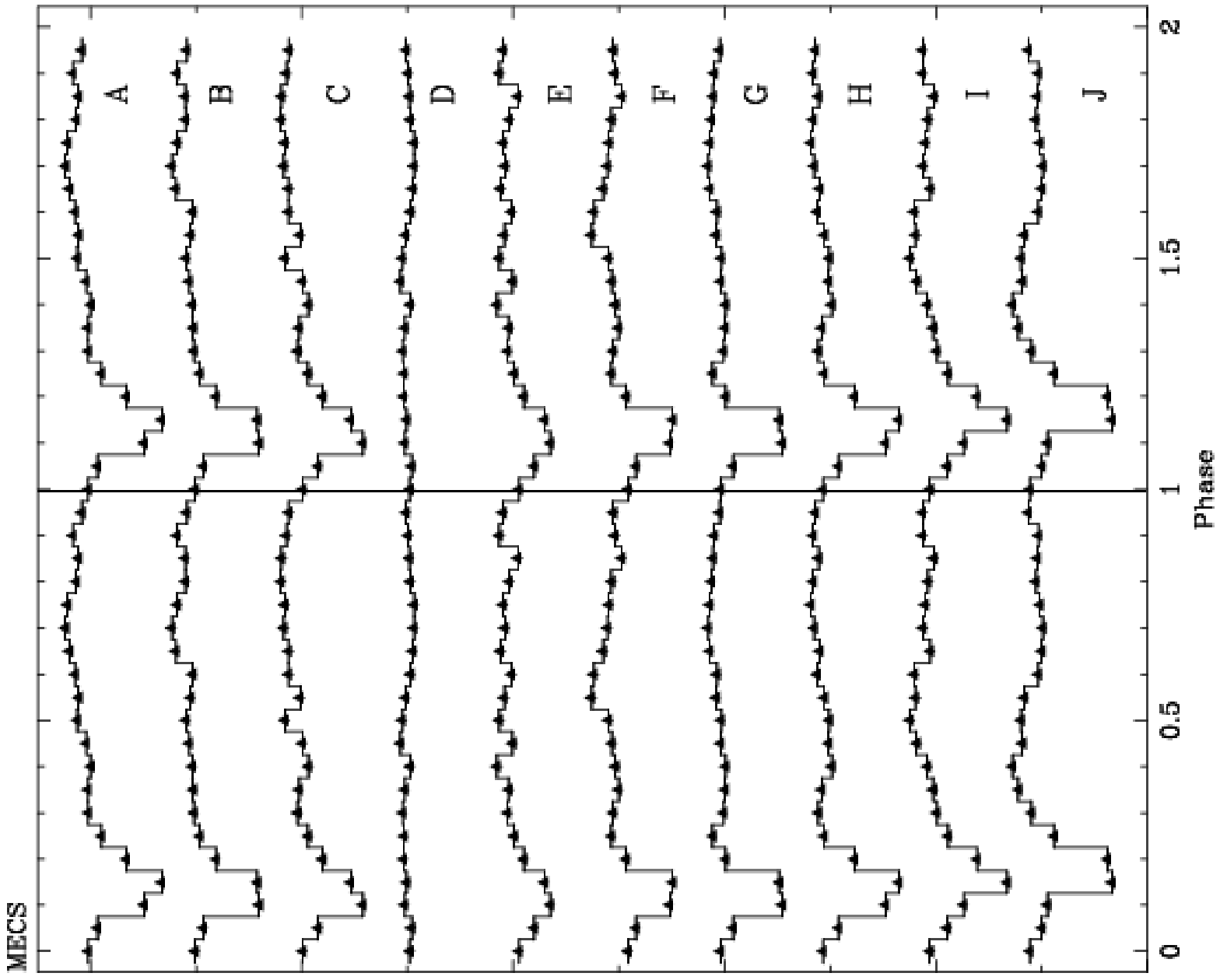,width=4cm,height=12cm,angle=270} 
\psfig{figure=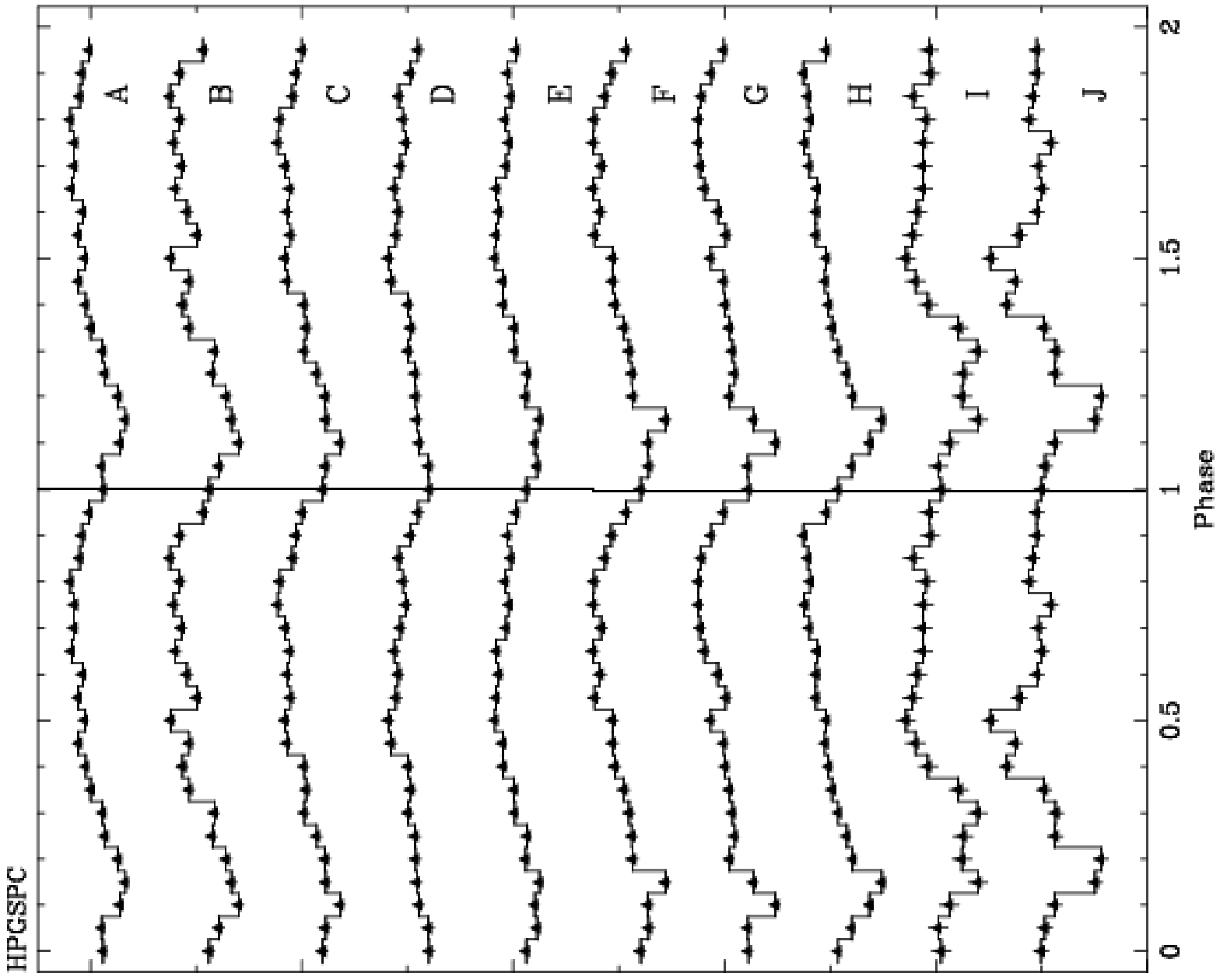,width=4cm,height=12cm,angle=270} 
\psfig{figure=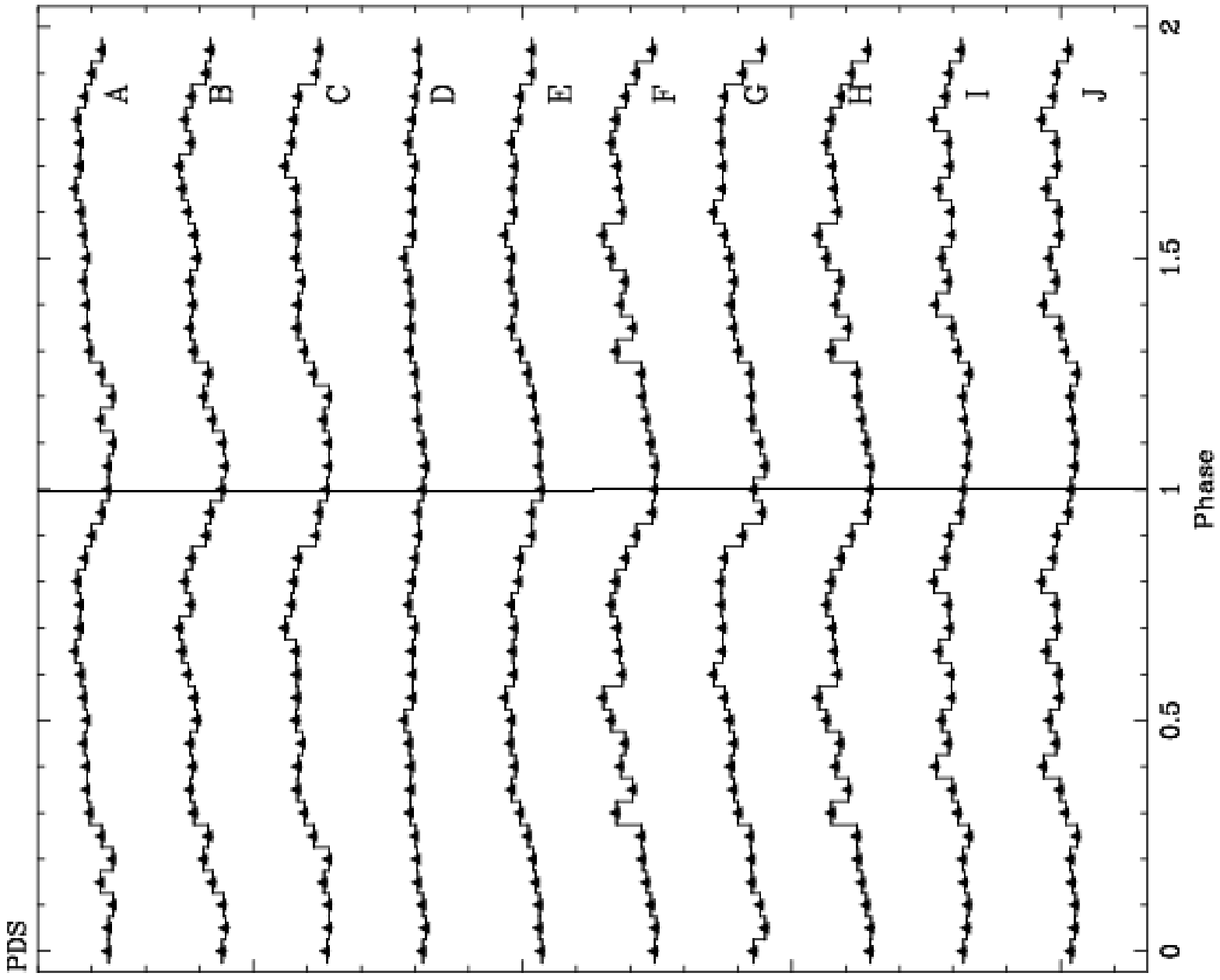,width=4cm,height=12cm,angle=270} }
\caption{LECS (0.2-4\,keV), MECS (1.65-7\,keV), HPGSPC (7-35\,keV) and PDS
(35-100\,keV) folded light-curves at the NS spin period for all time
intervals (see also Fig. \ref{lcurve_psd}). The scale on the y axis
are arbitrary normalised counts/s. The vertical lines are added in
order to help visualizing the shifts in the phase minima, more evident
in the third panel.}
\label{efold}
\end{figure*}
%%%%%%%%%%% FIGURE 3 %%%%%%%%%%%%%%%%%%%%%%%%

Pulsations were clearly seen in the Power Spectrum, the fundamental
frequency was followed by six harmonics (Fig.\,2 bottom
panel). We then carried out an Epoch Folding Search followed by
phase-fitting period determination, which gave a refined spin period
value of $P_{s} = 134.925\pm0.001$\,s (at 11785.000781 TJD; all errors
in the text, if not otherwise specified, are at 90\% confidence level
(c.l.); all errors in the figures are at 1$\sigma$ c.l.). The timing
analysis was carried out using {\em Xronos} tools version 5.19.

We analysed in the same way the previous two \BSAX observations (in
1996 and 1997) and derived a secular spin period derivative of
$\dot{P} = (1.0\pm0.2)\times10^{-7}$\,s\,s$^{-1}$ across the three
observations. Note that, although \gx \ shows a characteristic
spin-down trend, its timing behaviour has not been always stable, and
a period derivative reverse has been observed in a few occasions
(Chakrabarty et al. 1997b; Pereira, Braga \& Jablonski 1999).
Therefore, the period derivative estimated from the three \BSAX\,
datasets should be regarded as an average $\dot{P}$ value.

\begin{figure}
\psfig{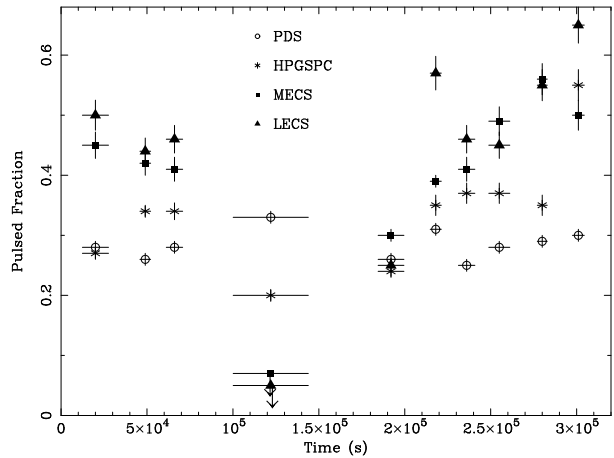} 
\caption{Pulsed Fractions for different instruments 
calculated with the formula: 
$ (I_{max}-I_{min})/(2I_{max}+ 2I_{min})$. Energy ranges for the four instruments are reported in the caption of Fig.\,3}
\label{pf}
\end{figure}

We divided the observation in 10 time intervals (Fig.\ref{lcurve_psd}
top panel), and searched for pulsations in each interval with all four
BeppoSAX instruments.  During the whole observation, but interval D,
all instruments showed pulsations at the same spin period. During the
search for pulsations in the low intensity state (interval D) we found
that in the LECS and in part of the MECS bands ($<$7\,keV) no pulsed
emission was present (upper limits on the pulsed fraction are 7\% and
5\% for MECS and LECS, respectively; see also Fig. \ref{pf}) while at
higher energies ($>$ 7\,keV) a quasi-sinusoidal pulsed signal was
always clearly detected (c.l. $>8\sigma$; see Fig. \ref{efold} and
\ref{pf}).

%%%%%%%%%%%%%%%%%%%%%% FIG. 4 %%%%%%%%%%%%%%%%%%%%%%%%

%\begin{center}
%\end{center}
%%%%%%%%%%%%%%%%%%%%%%%%%%%%%%%%%%%%%

In all four instrument we studied the pulse profiles in various energy
ranges, finding marked differences (see Fig.
\ref{efold}). By considering data taken with the 
same instrument and comparing the phase of the pulse minimum among the
HPGSPC and PDS pulse profiles, we found evidence for a shift in phase
between the pulses detected at different times. For instance, the
minima of the HPGSPC folded light-curves of intervals A and D are
shifted in phase by $0.22\pm0.05$ and there is evidence for a trend in
the phase shift evolution.
\footnote{ Because of the highly variable
pulse profile of this source, it is not possible to study
quantitatively the evolution of the phase shift during the
observation, since is rather difficult to find a common model that
fits the various light curves, and then an exact definition of the
pulse minimum. This issue can then be discussed only qualitatively.}
If we concentrate on the pulse minimum and we use interval D as a
reference, (see Fig.\ref{efold}, in particular panel three), the phase
shifts seems to decrease from interval A to D, and then to increase
again until the end of the observation.

Variations of the pulsed fraction (PF) are shown in Fig.\ref{pf}.
During the low intensity emission (interval D) the PF below 7\,keV is
consistent with zero (note that in Fig.\ref{pf} LECS and MECS pulsed
fractions in the interval D are 1$\sigma$ upper limits), while at
higher energies it increases to $20\pm4$\% (7-35\,keV) and $32\pm5$\%
(35-100\,keV). The PFs were inferred using the simple $
(I_{max}-I_{min})/(2I_{max}+2I_{min})$ formula; the high variability
of the pulse profile prevented us from using a more accurate
technique, like fitting with one or more sine functions.

\subsection{Spectral analysis}

Since \gx \ shows a large intensity variability, we considered
separately spectra corresponding to the high and low emission parts of
the light-curve studying the spectral changes over the 10 time
intervals used for the timing analysis (Fig.\ref{lcurve_psd} top
panel). 

We first fitted the spectra by using a simple absorbed cut-off power
law. This model gave a satisfactory reduced $\chi^2\sim 1$ only in the
intervals corresponding to a high source intensity (all but intervals
C, D, E and F, where the $\chi^2\sim 2$). We then added a blackbody
model in order to search for a soft component (present in other X-ray
binaries) but the addition of this further component was not
significant.

Concerning the spectra relative to the low intensity part of the
observation, we tried to fit them using the same broadband model
of the other spectra but adding a partial covering factor; the
resulting $\chi^2$ was about 1.5, better than the fit with the cut-off
power law alone but still quite large.

We then noticed that the source spectral shape in the low emission
state was reminiscent of that expected for a Compton reflection
dominated spectrum, we then fit all spectra by using an exponentially
cut-off power law (mimicking the incident X-ray beam) reflected by
some neutral material (which we tentatively associate to the disc;
pexrav model in {\em Xspec}; Magdziarz \& Zdziarski 1995).  By varying
the relative amount of reflected and primary components 
%(for matter subtending a 2$\pi$ solid angle at the primary source),
this provided an accurate model for all the \gx spectra taken in
different intensity states (see Fig.\,5 and Table 1). The reflection
parameter was small or even compatible with zero away from the low
intensity event, while during the latter it reached the highest values
($\sim 40$).

%%%%%%%%%%%%SPECTRA%%%%%%%%%%%%%%%%%%%%%%%%%

\begin{figure*}
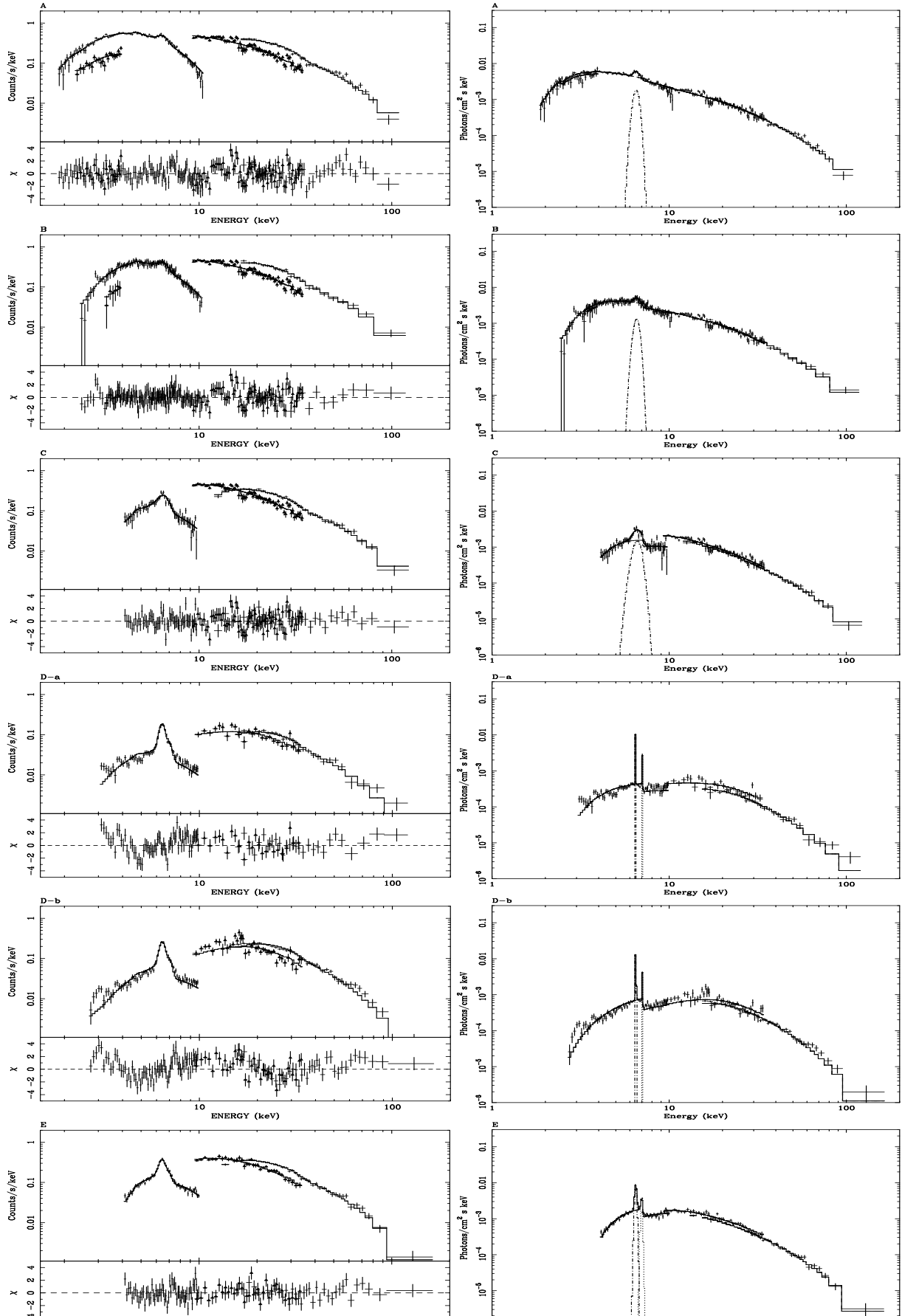

\caption{Time-resolved spectra. Time intervals are reported in Fig.\,1 and spectral parameters in Tab.\,1 .}

\hbox{
\vbox{
\psfig{figure=A_gen04.ps,width=8cm,height=4cm,angle=270} 
\psfig{figure=B_gen04.ps,width=8cm,height=4cm,angle=270} 
\psfig{figure=C_gen04.ps,width=8cm,height=4cm,angle=270} 
\psfig{figure=Da_gen04.ps,width=8cm,height=4cm,angle=270} 
\psfig{figure=Db_gen04.ps,width=8cm,height=4cm,angle=270} 
\psfig{figure=E_gen04.ps,width=8cm,height=4cm,angle=270} }
\vbox{
\psfig{figure=A_gen04_ufs.ps,width=8cm,height=4cm,angle=270} 
\psfig{figure=B_gen04_ufs.ps,width=8cm,height=4cm,angle=270} 
\psfig{figure=C_gen04_ufs.ps,width=8cm,height=4cm,angle=270} 
\psfig{figure=Da_gen04_ufs.ps,width=8cm,height=4cm,angle=270} 
\psfig{figure=Db_gen04_ufs.ps,width=8cm,height=4cm,angle=270} 
\psfig{figure=E_gen04_ufs.ps,width=8cm,height=4cm,angle=270} }
}
\label{spettri_uno}
\end{figure*}

\begin{figure*}
\hbox{
\vbox{
\psfig{figure=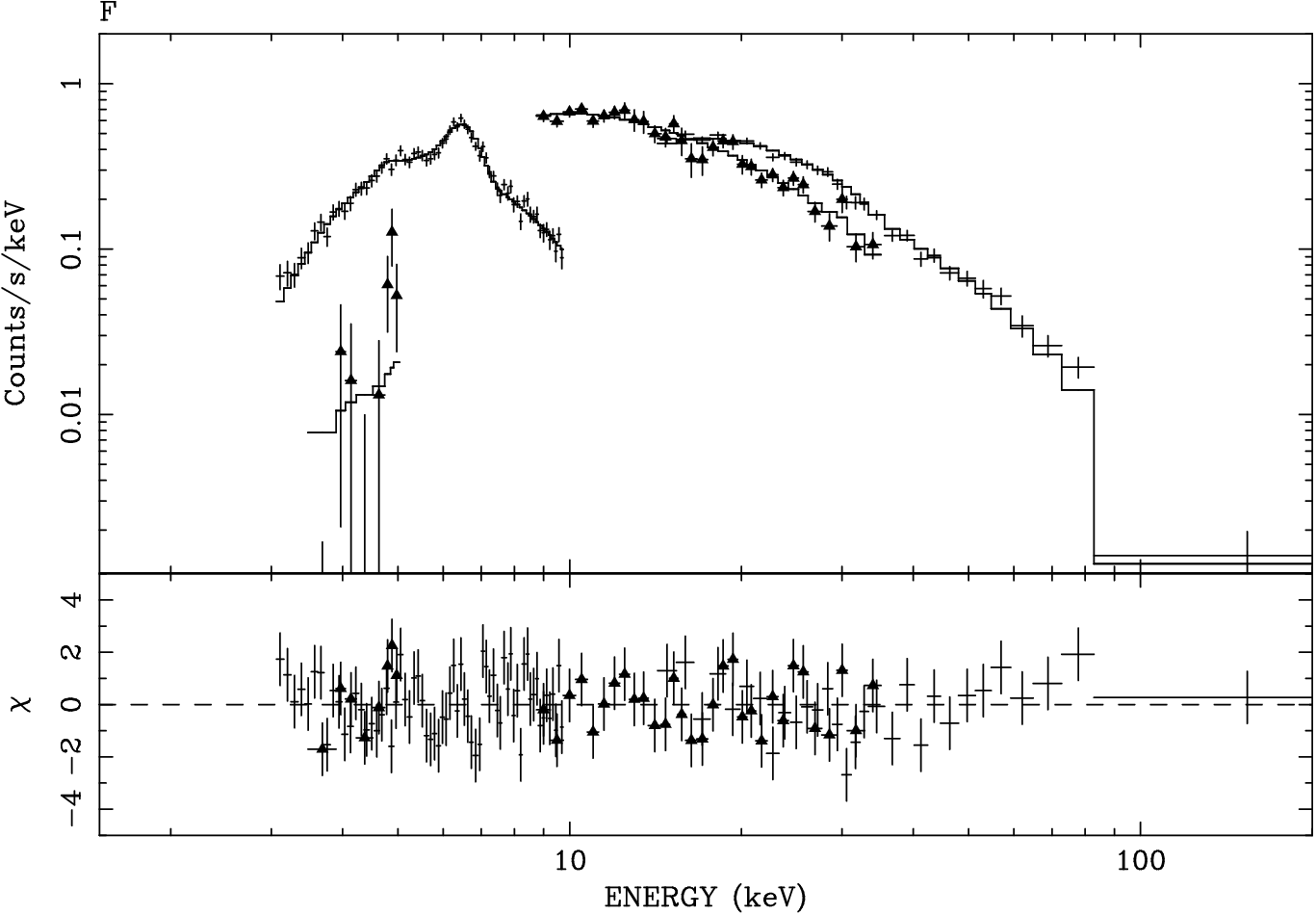,width=8cm,height=4cm,angle=270} 
\psfig{figure=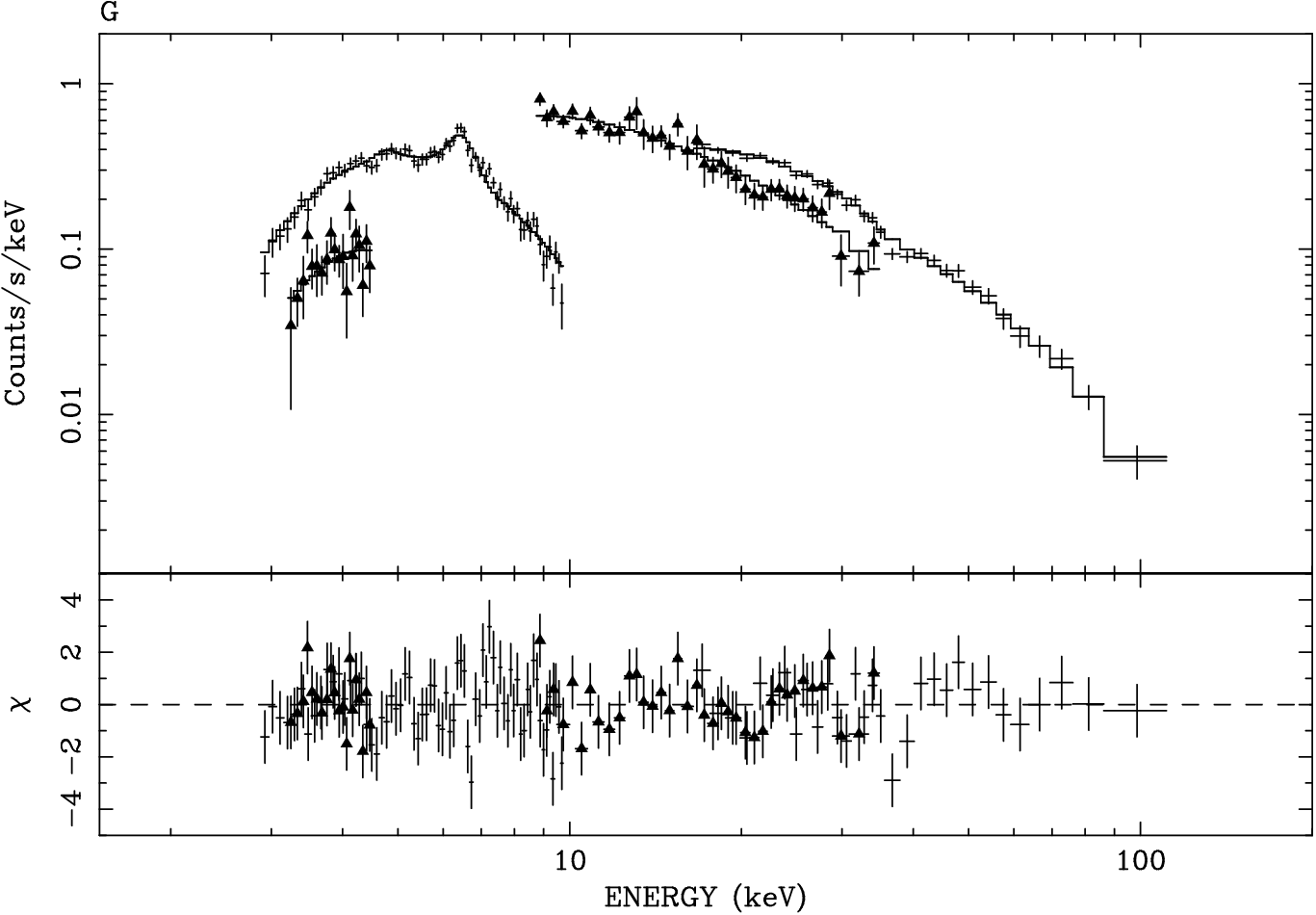,width=8cm,height=4cm,angle=270} 
\psfig{figure=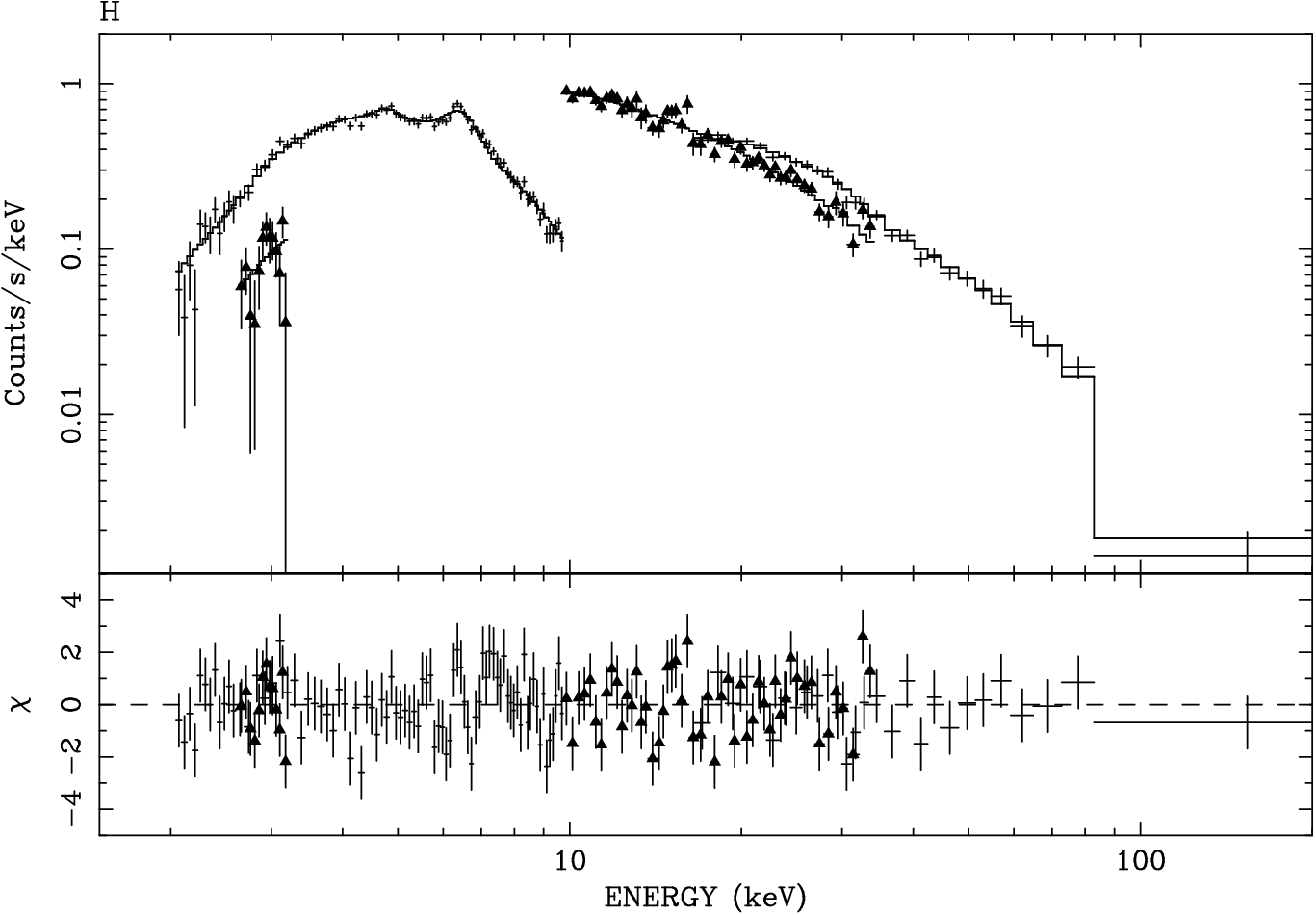,width=8cm,height=4cm,angle=270} 
\psfig{figure=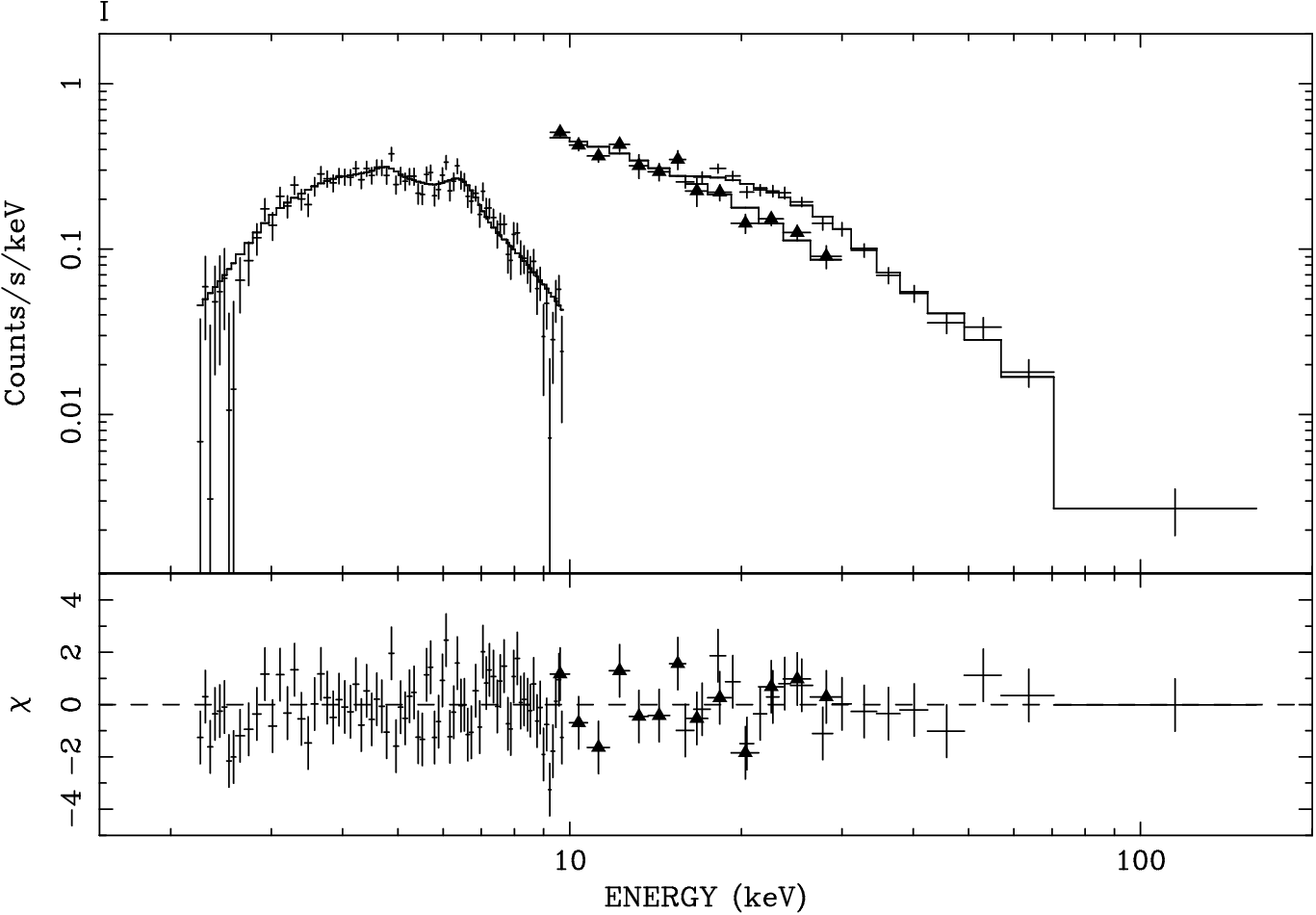,width=8cm,height=4cm,angle=270} 
\psfig{figure=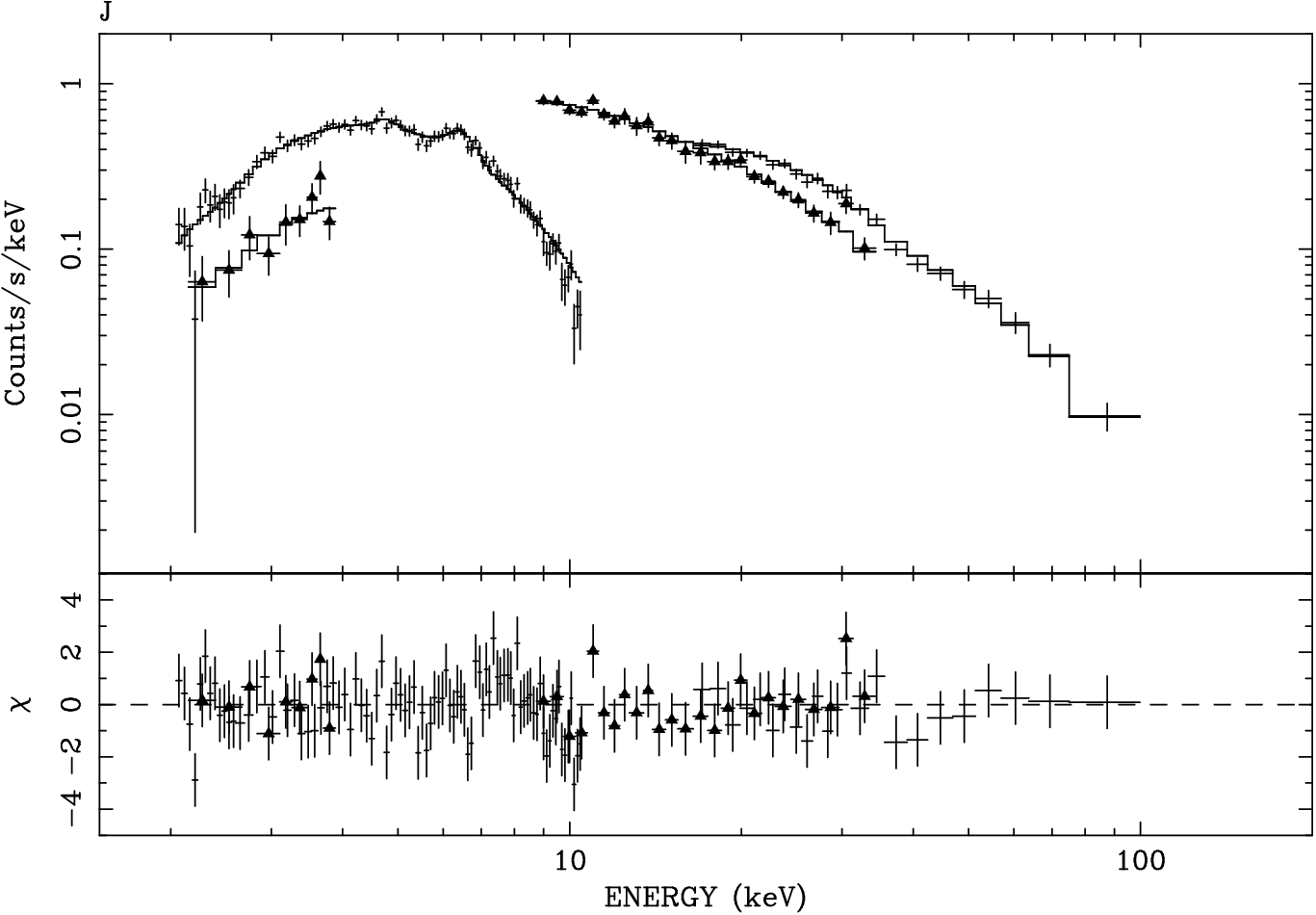,width=8cm,height=4cm,angle=270} }
\vbox{

\psfig{figure=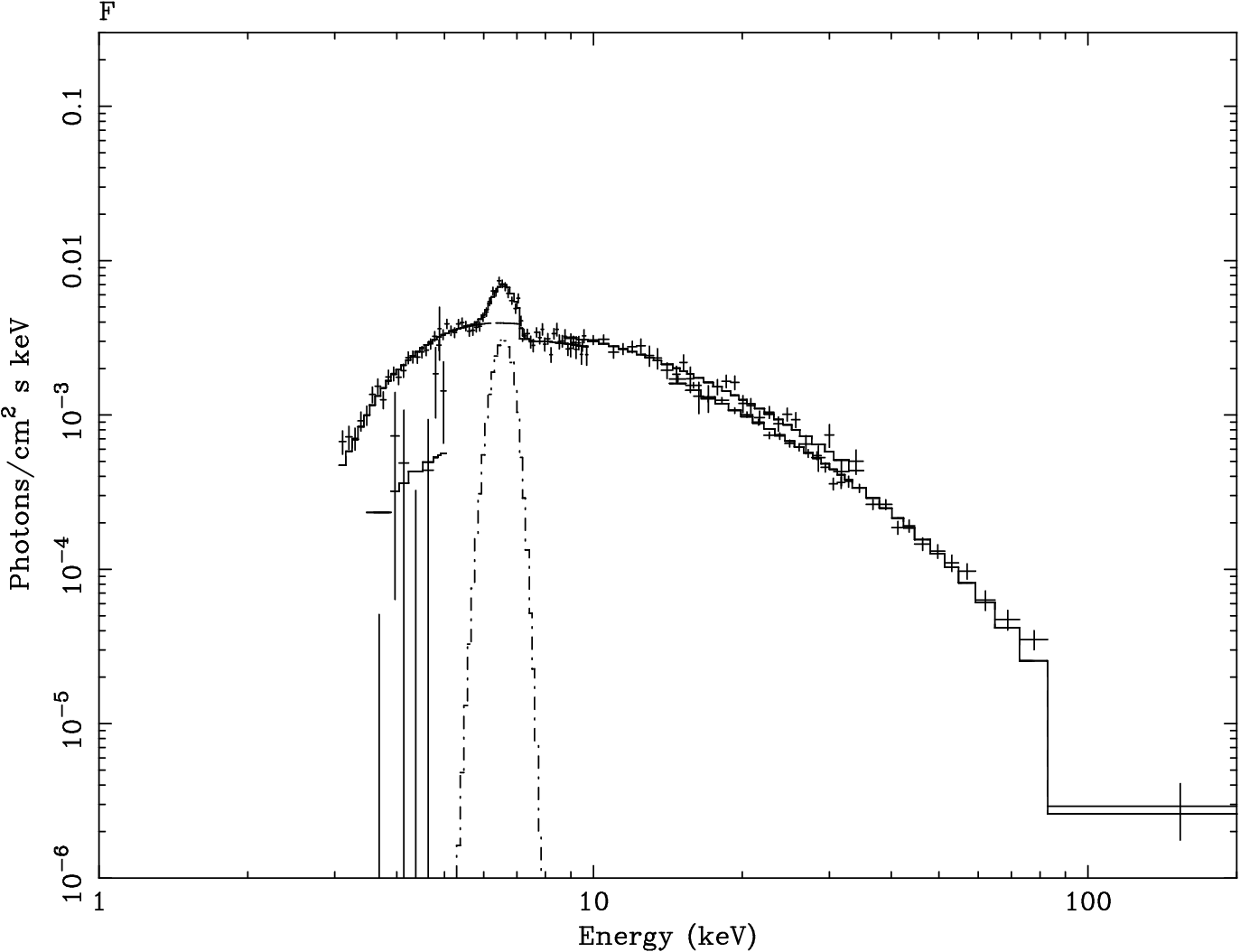,width=8cm,height=4cm,angle=270} 
\psfig{figure=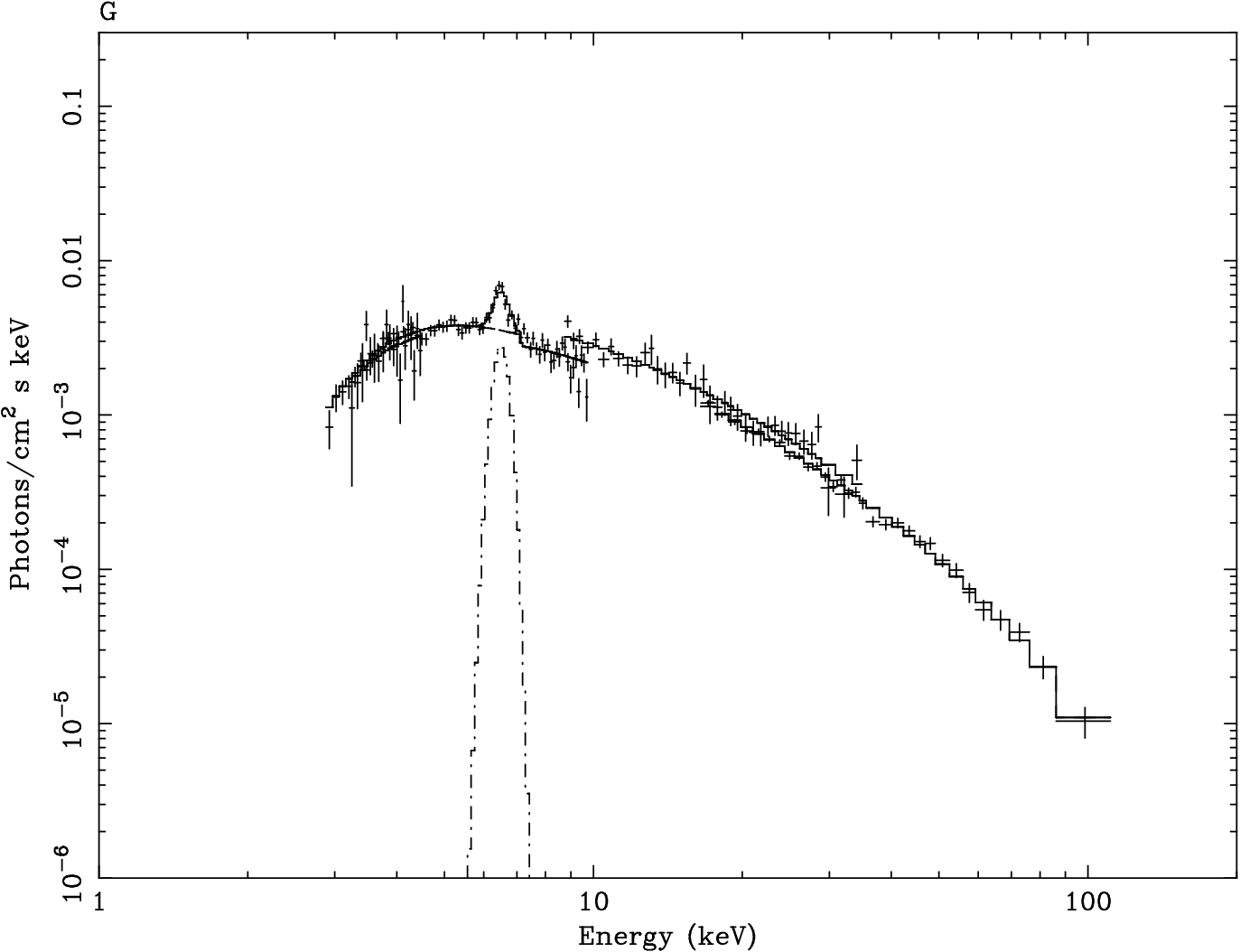,width=8cm,height=4cm,angle=270} 
\psfig{figure=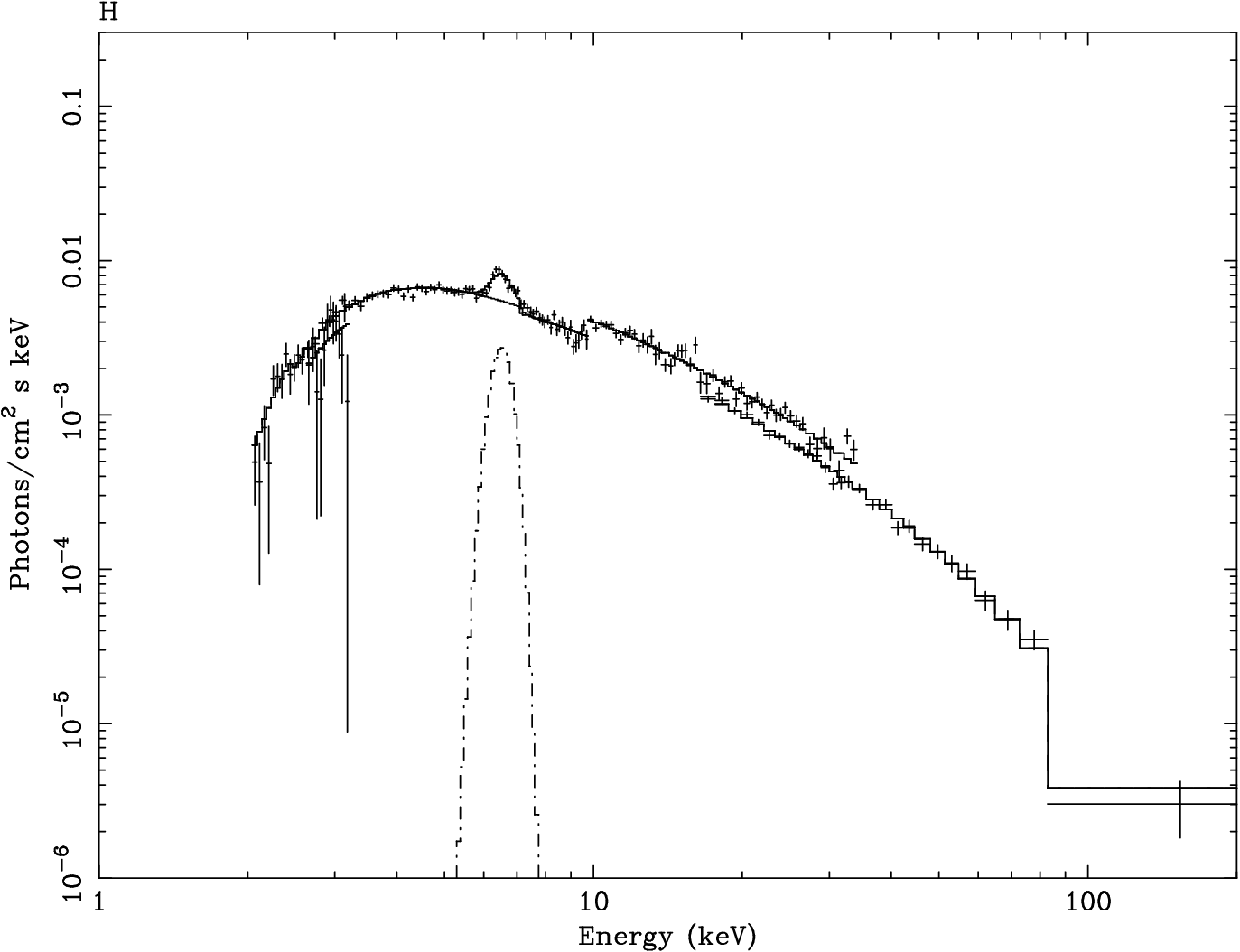,width=8cm,height=4cm,angle=270} 
\psfig{figure=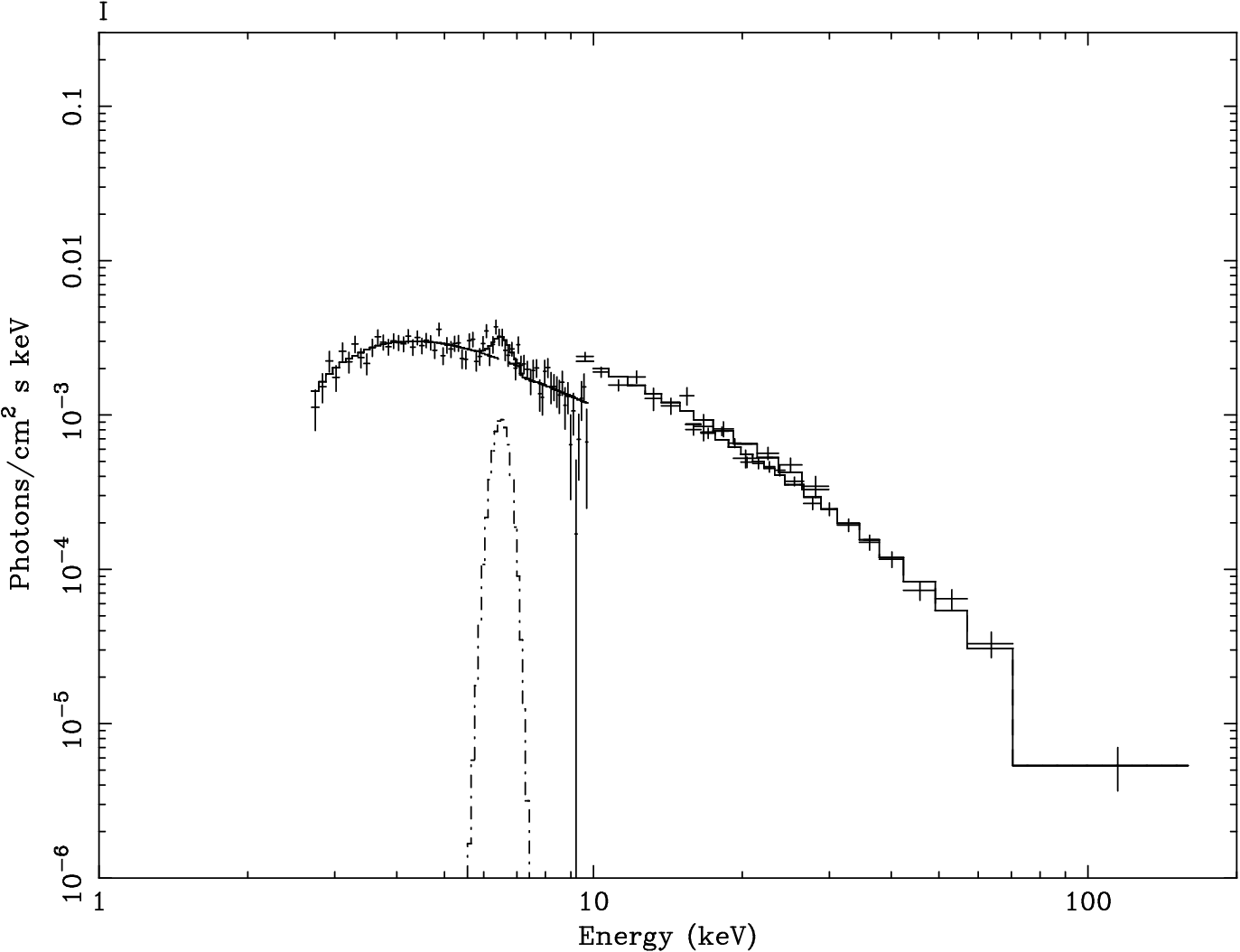,width=8cm,height=4cm,angle=270} 
\psfig{figure=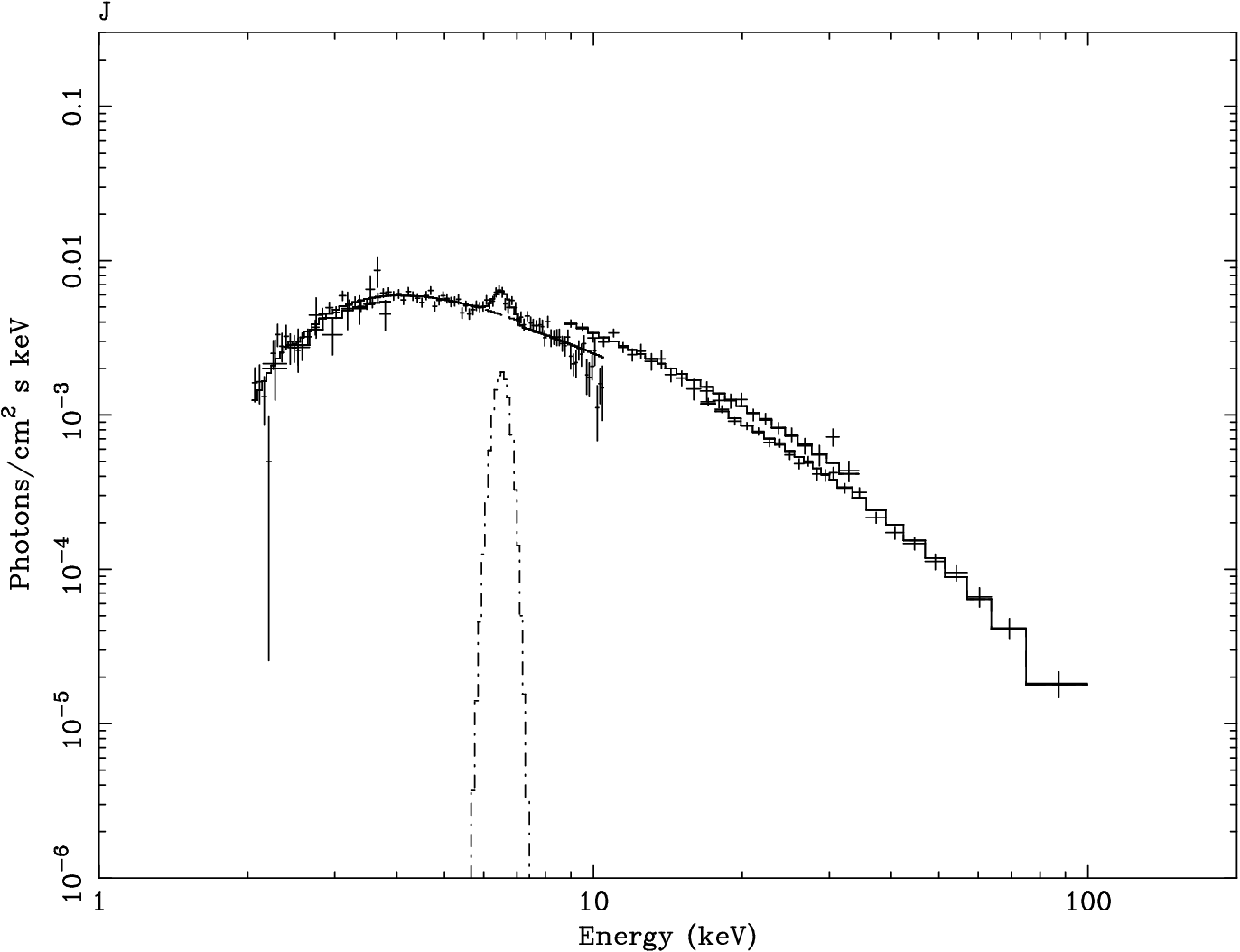,width=8cm,height=4cm,angle=270} }
}
\label{spettri_due}

\end{figure*}

%%%%%%%%%%%%%%%%%%%%%%%%%%%%%%%%%%%%%%%%%%

By adopting this spectral decomposition we found evidence also for
changes in the absorption parameter ($N_{H}$) and photon index
($\Gamma$) during the observation. The hydrogen column density $N_{H}$
varied between $7-48\times10^{22}$atoms\,cm$^{-2}$, reaching the
highest values just before and after the low intensity event. Note
that the average ISM absorption value in the direction of \gx
(computed using the nh tool provided by {\it heasarc}) is
$3.2\times10^{21}$atoms\,cm$^{-2}$, meaning that most of the
absorption is due to local material.

The power law index $\Gamma$ decreases by 10-20\% when the source
enter in the Compton reflection dominated phase (interval D) and the
spectrum remained hard until interval H, where a flaring event
occurred. This was followed by an abrupt decrease with spectral
softening, and then a recovering to a harder value.

%%%%%%%% TABLE %%%%%%%%%%%%%%%%%%%%%%%%%%%
\begin{center}
\begin{table*}
 \centering
 \begin{minipage}{126mm}
  \caption{Spectral parameters for different time intervals (intervals D-a and D-b are respectively the first and the second half of interval D).}
  \begin{tabular}{@{}lcccccc}
  \hline
  % Name     &            & \multicolumn{4}{c}{Flux density (Jy)%
  %\footnote{Observed by {\em IRAS}.}}\\
  & A & B & C & D-a  & D-b & E   \\
        
  \hline

&&&&&\\
 $ N_{H}(10^{22}$\,atoms\,cm$^{-2})$ & $7.4\pm0.2$ &   $17\pm1$  &    $39\pm5$ &   $22\pm2$        &   $20\pm6$ &   $48\pm3$   \\
&&&&&\\

Pexrav $\Gamma$             & $1.48\pm0.06$ & $1.7\pm0.1$  &$1.4\pm0.3 $ &  $1.3\pm0.1$ & $1.2\pm0.2$       & $0.9\pm0.2$\\
Pexrav $E_{fold}$ (keV)        & $42\pm3$ &  $57\pm12$   & $37\pm10$  &    $40\pm2$ &   $37\pm2$     &  $25\pm 4$ \\
Pexrav rel-refl               & $0.9\pm0.1$  & $1.2\pm0.3$  &  $1.2\pm0.8$  &   $8\pm2$     &   $42\pm3$    &    $0.14\pm0.08$  \\
Pexrav Norm  ($10^{-2}$)     & $7.9\pm0.8$ &  $ 12\pm3$ &   $5\pm2$   &  $0.5\pm0.1$   &   $0.2\pm0.1$ &      $3\pm1$ \\
& &&&&\\
 
$1^{st}$ Fe line (keV)        & $6.51\pm0.04$ & $6.54\pm0.07$  & $6.58\pm0.09$ &  $6.48\pm0.02$   &  $6.47\pm0.08$  & $6.47\pm0.05$  \\
$1^{st}$ line Width  (keV)      &$ 0.23\pm0.05$  & $0.2\pm0.1$  &  $0.3\pm0.1$   & $< 0.1$    &  $< 0.1$ &   $< 0.1$  \\
$1^{st}$ line Eqw (eV)          & $256$ &  $201$   & $860$  & $2160$  &  $1730$ & $823$\\
$1^{st}$ line Norm ($10^{-3}$)     & $1.18\pm0.21$ & $1.0\pm0.2$  &  $ 2.4\pm0.5$     & $1.3\pm0.2$   & $1.7\pm0.3$  & $3.2\pm0.1$   \\

$2^{nd}$ Fe line  (keV)        &  &   &   &   $7.05\pm0.11$    &    $7.05\pm0.11$    &   $7.05\pm0.18$\\
$2^{nd}$ line Width  (keV)          &  &   &   &   $< 0.1$    &   $< 0.1$     &  $< 0.1$ \\
$2^{nd}$ line Eqw (eV)          &  &  &   & $518$   &     $485$       & $ 201$ \\
$2^{nd}$ line Norm  ($10^{-3}$) & &&  &    $0.27\pm0.03$   &  $0.40\pm0.03$   &  $0.70\pm0.12$ \\

& &&&&\\

$ \chi^2/d.o.f.  $   &  1.1(221dof) & 1.2(225)    & 1.3(226) &  1.3(114)  & 1.3(141) &    1.2 (125) \\
Unab Flux  ($10^{-9}$erg\,cm$^{-2}$s$^{-1}$) &  $1.5$ &   $1.5$ & $0.87$ &  $0.35$  &   $0.59$ &    $1.3$\\

\hline
\hline
  & F & G & H & I & J & \\
\hline
&&&&&&\\
 $ N_{H}(10^{22}$\,atoms\,cm$^{-2})$ &   $ 25.2\pm1.4$ & $ 15\pm1$    & $ 9.7\pm0.3$  &  $ 11\pm1$    & $ 7.4\pm0.3$ &\\
&&&&&&\\

Pexrav $\Gamma$                & $0.9\pm0.1$    & $1.16\pm0.11$      &  $1.19\pm0.07$     &  $1.6\pm0.1$    &  $1.19\pm0.08$  &  \\
Pexrav $E_{fold}$ (keV)        &   $23\pm2$ & $30\pm3$    &   $31\pm1$       &    $41\pm5$    &   $30\pm2$     & \\
Pexrav rel-refl               &   $\sim 0$   &   $\sim 0$   &   $\sim 0$ &  $0.3\pm0.2$ &  $\sim 0$ & \\
Pexrav Norm   ($10^{-2}$)      & $4\pm1$ &  $5\pm1$   &  $7.2\pm0.8$    & $6\pm1$   & $5.6\pm0.7$ &\\
& &&&&&\\
 
$1^{st}$ Fe line (keV)     &  $6.55\pm0.03$    &  $6.50\pm0.03$ &   $6.51\pm0.03$     &  $6.50\pm0.03$  &  $6.52\pm0.03$   & \\
$1^{st}$ line Width  (keV)     & $0.32\pm0.05$ & $0.23\pm0.09$  &  $0.30\pm0.05$    & $0.25\pm0.07$    & $0.23\pm0.07$  & \\
$1^{st}$ line Eqw (eV)        &  $628$   & $421$   &  $379$   &  $263$ & $247$ & \\
$1^{st}$ line Norm  ($10^{-3}$)    &  $3.7\pm0.1$   &  $2.0\pm0.1$    &   $2.4\pm0.1$ &  $0.72\pm0.09$   &  $1.24\pm0.1$   &  \\

& &&&&&\\

$ \chi^2/d.o.f.  $    &   1.1(147)  &  1.2(166)  &   1.18(92)   & 1.08(138) &   1.05(185)  \\
Unab Flux  ($10^{-9}$erg\,cm$^{-2}$s$^{-1}$)  &  $1.9$ &  $1.9$ &   $1.9$ &  $0.76$ &   $1.5$ & \\

\hline
\end{tabular}
\end{minipage}
\end{table*}

\end{center}
%%%%%%%%%%%%%%%%%%%%%%%%%%%%%%%%%%%%%%%%%

All time-resolved spectra showed at least a Fe line emission feature.
In all intervals but D and E, only one broad ($\sim$0.3\,keV) Fe line
emission was present at $\sim$6.55\,keV. In several other sources a
broad Fe line was detected around this energy, but then observations
with higher spectral resolution instruments ({\em Chandra}-HETG and
XMM-Newton) disentangled the blending of two narrow lines at $\sim$6.4
and 6.7\,keV (Audley 1997; Gallo et al.  2004).  In order to
investigate the possible occurrence of blending, we fitted the spectra
adding to the continuum model two Gaussians with fixed peak energies
at 6.4 and 6.7\,keV (neutral and He-like Fe $K_{\alpha}$,
respectively), and forcing their widths to be equal. The fit gave
narrow widths ($<$0.1\,keV, which is the energy resolution of the
instrument at that energy) and a ratio between the two line
normalisations was variable in time ranging between $I_{(6.7\,keV} /
I_{6.4\,keV} \sim 0.2-0.6$. However, the reduced chi-squared remained
close to one with three more parameters and the F-test revealed the
addition of a new Gaussian function not significant. Only high
resolution observations could shed more light on this topic.

In intervals D and E, while the continuum reduced substantially, an
additional narrow line at $\sim$7.05\,keV was clearly revealed. At the
same time, the broad line at $\sim$6.55\,keV became narrower and its
centroid shifted at slightly lower energies.  The $\sim 7$keV line was
possibly present also in higher intensity spectra but being too weak
to be detected because of the intense continuum emission.  A natural
interpretation is that the broad line detected during the high
emission results from the blending of a 6.4\,keV neutral Fe
$K_{\alpha}$ and a 6.7\,keV He-like Fe, while the two narrow lines in
the low intensity emission are the neutral Fe K$\alpha$ and
K$\beta$. The K$\beta$ flux is about 20\% that of the K$\alpha$,
slighly larger than the expected value of 15-16\% (Molendi et
al. 2003).  However, we would like to stress the fact that the
relatively poor MECS energy resolution makes our interpretation of
these lines quite uncertain. For example, we cannot exclude that the
line at $7.05\pm0.11$\,keV can be, at least in part, due to the
6.93\,keV Fe XXVI line.

In some spectra (especially during intervals A and G, there was a weak
evidence for an absorption feature in the 32-37\,keV range, similar to
an edge or a cyclotron line. An absorption edge at this energies is
unlikely, while fitting a cyclotron model (cyclabs in {\em Xspec}) we
found an improvement of the chi-square value with an F-test
probability of $2.5\sigma$ and $2.3\sigma$ for interval A and G,
respectively.  If real, the feature can be interpreted as an electron
cyclotron feature. The inferred surface magnetic field of the NS would
then be $\sim4\times10^{12}$\,G, much lower than the value proposed by
interpreting the torque reversal of the source (Dotani et al. 1989).
In order to better study the possible presence of a cyclotron line, we
made a Pulse Phase Spectroscopy (PPS) analysis in time-intervals A and
G in the whole 1--200\,keV energy band. In fact cyclotron lines are
often expected to have a spin-phase dependent strength, but
unfortunately the PPS analysis did not yield any improvement of the
line significance.  Further observations (e.g. INTEGRAL) are needed to
confirm this very weak evidence.

\section{DISCUSSION}

We reported on the results of a long \BSAX\ observation of the X-ray
binary pulsar \gx, during which the source entered a phase of low
X-ray emission. During the low event: i) a Compton reflection
dominated spectrum was clearly detected, for the first time in this
source, ii) pulsations were not detectable with the low energy
instruments, while they were clearly visible above $7\,$keV and iii)
an emission line at $\sim$7.05\,keV was revealed for the first time in
this source.

Moreover we found that the pulse profile of \gx \ is highly energy and
time dependent, and the pulse minimum at the higher energies
systematically shifted in phase during the whole observation.  A broad
Fe line at $\sim$6.5\,keV, detected also with past missions, was
present in all spectra while the substantial reduction of the
continuum emission during the low intensity event, allowed us to
reveal, for the first time, a second Fe line at $\sim$7.05\,keV.

A similar (although shorter) low emission event was observed in \gx\
with RXTE (Giles et al. 2000; Galloway et al. 2000), but due to the
poor RXTE spectral resolution a comparison between the spectra of
these two events is not straightforward. During a few occasions \gx
was detected by RXTE in an unusual low flux non-pulsating state (Cui
1997; Cui \& Smith 2004).  The source fluxes in such observations were
comparable to the flux of the source during the low intensity event
reported here. Cui (1997) also pointed out that the X-ray spectrum was
harder than usual and interpreted these non-pulsating events as due to
an onset of a centrifugal barrier during the transition to the
propeller regime.  However, although such events were extremely
similar to the one reported here, an interpretation in terms of the
onset of centrifugal barrier appears less likely since no pulsations
are expected in the propeller regime, if accretion onto the magnetic
poles is halted, neither at low nor at high energies. There could also
be particular magnetospheric configurations through which, even in the
propeller regime, matter might reach the magnetic poles of the NS
causing pulsations (see also Campana et al. 2001). However, even in
this case, it is not obvious to explain why pulsations are visible
only at high X-ray energies and the pulse minima shift in phase with
time.

In order to interpret the low intensity event occurred during the \BSAX\
observation, we present below different models we have examined.

A crucial insight derived from the fact that we found a common
spectral model that describes spectra taken at all the observed
intensity states, merely by varying the relative contribution of the
reflected component. The onset of a Compton reflection component is
unlikely to be due to a different emission state of the source
itself. It is instead most probably associated with the reprocessing
of the NS emission by some Compton thick material (possibly the
accretion disc that partially intercepts the NS X-ray beam).
Moreover, the high variability in the absorption value detected during
the \BSAX \ observation and the lack of X-ray pulsations at low
energies during the low event, do not find a natural explanation in a
scenario were the onset of the low state is caused by a variation of
emission of the NS itself.

Another possibility is a variation in the accretion rate from the
companion star (as proposed by Galloway et al. 2000 for the other low
emission event). While variations might well be present, they cannot
explain alone the onset of a pronounced reflection component, the
highly variable absorption, the unpulsed emission at low energies or
the varying shift in phase of the high-energy folded lightcurves.

Our spectral analysis indicates that the spectrum observed during the
low intensity event should not be due to the direct NS surface
emission. Rather most of the photons we see, reach us after reflection
off material located in the proximity of the NS. This material might
be due to the disc surrounding the NS or to the high density wind of
the companion, perhaps accompanied by a sort of eclipse from the giant
companion itself.  Below we discuss these possibilities.

%%%%%%%%%%%%%%%%%%%%%%%%%%%%%%%%%%%%%%%
%\begin{center}
\begin{figure}
\psfig{figure=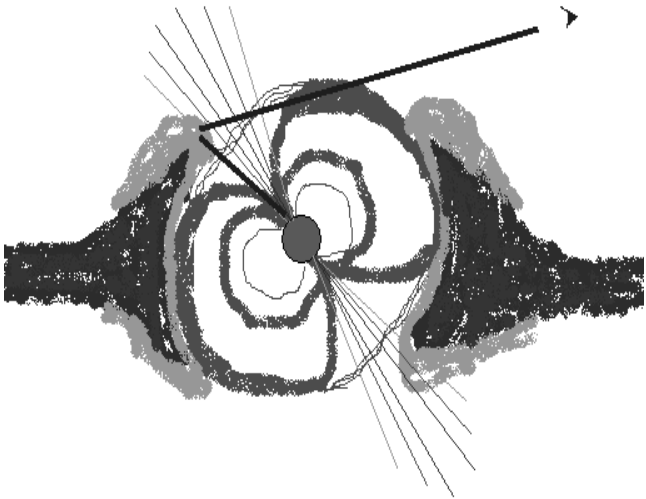,width=8.5cm,height=8cm}
\caption{Schematic two dimensional picture of the torus--like
accretion disc around the NS. Before the increase of the accretion
rate the torus was smaller and the observer could see the direct NS
emission, as the torus start to thicken, the direct emission is hidden
while the reflected component (solid line) prevails.}
\label{model}
\end{figure}
%\end{center}
%%%%%%%%%%%%%%%%%%%%%%%%%%%%%%%%%%%%%

The observed decrease in X-ray intensity might be due to a partial
eclipse, for instance the direct NS emission might be hidden, from our
line of sight, by the limb of the giant companion. In this case, the
X-ray spectrum observed during the eclipse would be mainly produced
through the Compton reflection off the material surrounding the
companion, e.g. its dense stellar wind.

The observed changes in $N_{H}$ are naturally explained by this
scenario. Since the stellar wind is denser and denser approaching the
giant surface, $N_{H}$ is expected to reach the highest values soon
before the eclipse ingress (interval C) and soon after the eclipse
egress (interval E), because at these epochs the direct X-ray photons
pass through a more dense material before their detection.

Furthermore, if the emission lines observed with \BSAX \ would be a
blending of the 6.4 and 6.7\,keV Fe during the high-intensity emission
(respectively neutral and ionized Fe; see also section 3.2) and the
6.4 and 7.0\,keV features would be neutral Fe lines during the low
emission event, this could be in agreement with the
eclipse scenario. In fact, during the partial eclipse the regions
closer to the NS, which are expected to be the most ionized, are
covered by the companion.  We then expect to detected ionization lines
when the NS is not obscured and lines from neutral material during the
eclipse, when only the farther regions are visible.

However, in order to produce such highly reflection dominated
spectrum, this model requires that the Compton thick material is
distributed along a wide solid angle around the source, and this
distribution is difficult to be ascribed by stellar companion wind
material only. Moreover, taking in account the large size of the
companion star compared with the NS radius, the occurrence of a
$\sim$90\,ks eclipse requires an ad hoc fine tuning of the system
inclination. Both problems might be partially reduced if we assume
that at least part of the scattering material is located in the
accretion disc. For certain inclination angles, the disc could subtend
a wide solid angle around the NS, and because of the large disc size,
compared to the NS radius, a lesser degree of fine tuning would be
required. However, this scenario requires a different tuning, now on
the inclination of the disc itself.

Another possible scenario, the one we favor, is one in which the
material responsible for the occultation of the direct NS emission
might be provided by a torus-like accretion disc around the compact
source (see Fig.\ref{model}). The increase in the accretion rate from
the companion star, possibly due to a change in the wind parameters,
would cause the inner torus to thicken\footnote{Note that the
occurrence of accretion rate changes in the history of this source is
well supported by the detections of spin-torque variations in the
pulse timing history (Chakrabarty et al. 1997b).}, thus hiding the
direct emission of the NS from our line of sight.

In this scenario the Compton reflected emission observed during the
low event is well explained in terms of emission reflected by the side
of the torus facing our line of sight.

In order to prevent light-travel time smearing of the pulsations, the
reflecting zone should not be farther than
$cP_{s}\sim4\times10^{12}$\,cm (where c is the light velocity and
$P_s$ the source spin period). The inner radius of the accretion disc
is dictated by the magnetospheric radius ($R_m$) of the NS. Since
torque reversals has been detected in \gx \, $R_m$ must be close to
the corotation radius, $R_{m}\sim
R_{cor}=(GM_{x}P^2_{s}/4\pi^2)^{1/3}=8.2\times10^9$\,cm, which is
smaller than $cP_{s}$.  Furthermore, for a $P_{s}$=134.925(1)\,s
pulsar a phase shift between pulse minima of $\sim 0.22$, as that
observed across intervals A and D (see section 3.1), corresponds to a
delay time of $29\pm6$\,s and to a difference in photon path of
$\sim8.7\times10^{11}$\,cm. In this picture, the lack of low energy
pulsations may be explained if we consider a reflecting material
composed of different layers, the near at the NS the Compton thicker,
the farther the Compton thiner. In fact if the Compton thiner material
is expected to be mainly responsable for the low-energy reflection, it
could be too far from the source to keep the coherence in the
low-energy photons that are reflecting.

Moreover, the $N_{H}$ variation can be understood as follow: by
increasing the torus the amount of materials in the line of sight
starts to increase, and so the $N_{H}$ value (intervals B and C). When
the direct emission is completely hidden (interval D) the $N_{H}$
value decreases because only the reflected component is seen. The
egress from the low state is an epoch when the torus starts to shrink
down again (intervals E and F) and the direct emission slowly re enter
into view, in the beginning through a large amount of material.

This scenario strengthes the interpretation of the two narrow lines
present during the event as the neutral Fe $K_\alpha$ and $K_\beta$
and the broad line as the blending of the neutral Fe $K_\alpha$ and
the He--like Fe $K_\alpha$. In fact, since out of the low intensity
event the whole disc emission is visible (in particular the disc
regions close to the NS which are highly ionised by the intense NS
X-ray beam) we then expect to see both neutral and ionised Fe emission
lines. On the other hand, when the torus hides the NS from the direct
view, only the reflection by the external regions is visible, we then
expect to detect only neutral lines.

The intensity of the neutral iron line is expected to correlate with
the intensity of the reflection component (e.g. Matt et al.  1991,
George \& Fabian 1991).  Unfortunately, the poor BeppoSAX energy
resolution makes difficult to separate the neutral and ionized
lines. Nevertheless, we tried to fit the iron line with two unresolved
(i.e. $\delta$-functions) lines with energies fixed to 6.4 and 6.7 keV
(the neutral and He-like ions, respectively).  The fit is as good as
the one with the broad line. The EW of the neutral line is very large
(of the order of 1-2 keV) when the reflection component dominates the
spectrum, as in interval D, as expected from MonteCarlo simulations
(e.g. Matt et al. 1991). In intervals A-C the line EW is much smaller
(up to about 250 eV), because of the dilution by the direct power law.
In intervals F-J, when the reflection component is very small, there
is still a substantial neutral line (EW of about 100-200 eV), which
can be due to the absorbing matter, which has a large covering
factor (e.g. Matt 2002).  Interval E is instead puzzling, as there is
still a very large neutral iron line (EW$\sim$800 eV) and a small
reflection component. In the framework of our proposed scenario,
however, it is possible that the torus, while shrinking, becames also
less optically thick (or at least does so the visible part of it). In
fact, for Thomson optical depths of a few tenths a still very intense
iron line is expected, while the reflection continuum is much reduced,
especially above 10 keV (Matt et al. 2003).

The model described above is analogous to that proposed for some AGN
(Walter et al. 2003; Revnivtsev et al. 2003; Matt \& Guainazzi 2003).

The $N_H$ value observed in \gx during some part of the observation is
similar to that measured in some of the highly absorbed INTEGRAL
sources, we could speculate that the highly absorbed INTEGRAL sources
are compact binary systems (as \gx) where, due to a particular line of
sight and inclination, the direct emission of the compact object does
not reach the observer being always hidden by the inner accretion
torus.

While this work was being drafted, Naik et al. (2004) published a
paper on the same observation. Their spectral modeling is not fully
consistent with ours. Naik et al. claimed the presence of a soft
excess that we do not detect. One possibility could be the different
technique they adopted for the background extraction. The technique
they used may led to underestimate the background at energies
$\leq$2\,keV, then a low-energy excess might have resulted (see Sec. 2
and 3 for further explanation). Moreover, they fit the spectrum of the
low intensity event with a Comptonization model with a higher
temperature of seed photons and a higher absorption value than in the
normal emission state. However, it is not clear how a hot Comptonizing
material might produce a 2.1\,keV equivalent width neutral Fe emission
line, as present during this low intensity event. These authors did
not use the HPGSPC instrument in the analysis, which gives a large
help in constraining the broad band spectrum, furthermore they divided
the observation in three time intervals, which we believe not
enough to reveal and model the whole spectral changes occurred during
the observation. \\

N.R. acknowledges Andrzej Zdzriaski, in charge as the referee of the
COSPAR proceedings, and the referee for several key comments.
N.R. thanks A. Antonelli, F. Fiore and T. Mineo for their advices on
the \BSAX\, analysis, F.Verrecchia for the analysis of the Wide Field
Camera observations and F. Mirabel for enlightening questions during a
seminar in Saclay where this work was presented. N.R. also thanks
L. Burderi, T. Di Salvo and M. M\'endez for useful discussions. This
work was partially supported by ASI and MIUR grants. N.R. is supported
by a Marie Curie Fellowship to NOVA.

\end{document}